\documentclass[a4paper,11pt]{article}

\usepackage[numbers]{natbib} 

\usepackage{authblk}
\usepackage{pdfpages}

\usepackage{array,multirow,graphicx}
\usepackage{rotating, float} 

\usepackage{color, colortbl}
\usepackage{hyperref}
\usepackage{xcolor,lipsum}
\usepackage{titlesec}
\definecolor{darkblue}{rgb}{0.12,0.47,0.87}
\definecolor{ceruleanblue}{rgb}{0.16, 0.32, 0.75}
\definecolor{white}{rgb}{1.0, 1.0, 1.0}

\titleformat{name=\section}[block]
  {\sffamily\Large}
  {}
  {0pt}
  {\colorsection}
    {\Large}
\titlespacing*{\section}{0pt}{\baselineskip}{\baselineskip}

\newcommand{\colorsection}[1]{%
  \colorbox{ceruleanblue!90}{\parbox{\dimexpr\textwidth-2\fboxsep}{\thesection\ #1}}}
  
 \titleformat{name=\subsection}[block]
 {\sffamily\large}
  {}
  {0pt}
  {\colorsubsection}
\titlespacing*{\subsection}{0pt}{\baselineskip}{\baselineskip}

\newcommand{\colorsubsection}[1]{%
  \colorbox{ceruleanblue!20}{\parbox{\dimexpr\textwidth-2\fboxsep}{\thesubsection\ #1}}}

\definecolor{LightCyan}{rgb}{0.53,0.81,0.98}
\definecolor{Gold}{rgb}{1,0.84,0}
\definecolor{Pink}{rgb}{1,0.75,0.80}
\definecolor{Green}{rgb}{0.6,0.98,0.60}
\definecolor{LightGreen}{rgb}{0,1,0}
\definecolor{Burlywood}{rgb}{0.87,0.72,0.53}
\definecolor{Snow2}{rgb}{0.93,0.91,0.91}

\definecolor{beaublue}{rgb}{1.0,1.0,1.0}

\definecolor{LightCyan}{rgb}{1.0,1.0,1.0}
\definecolor{Gold}{rgb}{1.0,1.0,1.0}
\definecolor{Pink}{rgb}{1.0,1.0,1.0}
\definecolor{Green}{rgb}{1.0,1.0,1.0}
\definecolor{LightGreen}{rgb}{1.0,1.0,1.0}
\definecolor{Burlywood}{rgb}{1.0,1.0,1.0}

\definecolor{Snow2}{rgb}{0.93,0.91,0.91}
\definecolor{beaublue}{rgb}{0.74,0.83,0.90}
\definecolor{blue(pigment)}{rgb}{0.2, 0.2, 0.6}
\definecolor{antiquefuchsia}{rgb}{0.57, 0.36, 0.51}
\definecolor{brickred}{rgb}{0.8, 0.25, 0.33}
\definecolor{brightmaroon}{rgb}{0.76, 0.13, 0.28}

\def\AND{$^,$}
\def\INSTIAPS{$^{1}$}
\def\INSTUNIBERN{$^{2}$}
\def\INSTGRENOBLE{$^{3}$}
\def\INSTDAN{$^{4}$} 
\def\INSTTUBS{$^{5}$}
\def\INSTINAFFI{$^{6}$}
\def\INSTJPL{$^{7}$}
\def\INSTLISA{$^{8}$}
\def\INSTCSHBERN{$^{9}$} 
\def\INSTCSNSM{$^{10}$}
\def\INSTICL{$^{11}$}
\def\INSTMPS{$^{12}$}
\def\INSTLPC2E{$^{13}$} 
\def\INSTOCA{$^{14}$} 
\def\INSTRIESTE{$^{15}$}  
\def\INSTESO{$^{16}$} 
\def\INSTLATMOS{$^{17}$} 
\def\INSTSWRI{$^{18}$} 
\def\INSTIGPCH{$^{19}$} 
\def\INSTROE{$^{20}$}
\def\INSTDLR{$^{21}$}
\def\INSTDLRBERLIN{$^{22}$}

\begin{document}


\title{{\bf{\Huge \textcolor{blue(pigment)}{AMBITION} \\ \LARGE \textcolor{blue(pigment)}{Comet Nucleus Cryogenic Sample Return}}}\\
{\normalsize {\bf A white paper for ESA Voyage 2050 long-term plan}}}

\author{}
\date{}
\maketitle


\begin{center}
{\bf Dominique Bockel\'ee-Morvan} \\
LESIA, Observatory of Paris, PSL University, CNRS, Sorbonne University, University of Paris\\
5 place Jules Janssen, 92195, Meudon Cedex, France\\
email: dominique.bockelee@obspm.fr \\
\vspace{+0.5cm}
{\it July 2019}
\end{center}

\vspace{0cm}

\begin{center}
\includegraphics[angle=0,width=10cm]{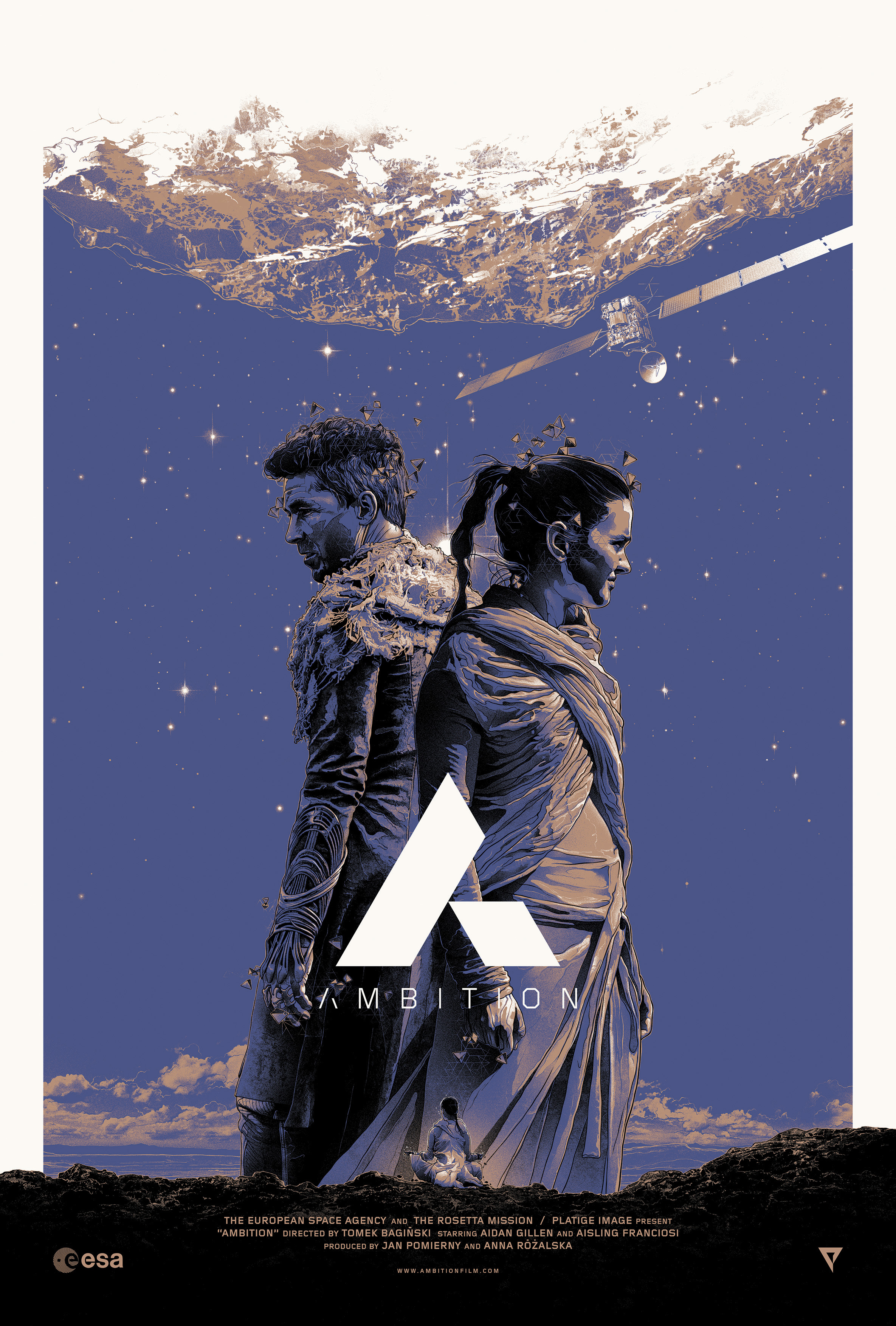}
\end{center}

\begin{center}
AMBITION film poster. Credit: ESA \& Platige Image.
\end{center}

\clearpage

\noindent
{\bf {\Large Team members}}

\noindent

\noindent
Gianrico Filacchione\INSTIAPS,
Kathrin Altwegg\INSTUNIBERN, Eleonora Bianchi\INSTGRENOBLE, Martin Bizzarro\INSTDAN, Ju\"rgen Blum\INSTTUBS, Lydie Bonal\INSTGRENOBLE, Fabrizio Capaccioni\INSTIAPS, Claudio Codella\INSTINAFFI\AND\INSTGRENOBLE, Mathieu Choukroun\INSTJPL, Herv\'e Cottin\INSTLISA, Bj\"orn Davidsson\INSTJPL, Maria Cristina De Sanctis\INSTIAPS, Maria Drozdovskaya\INSTCSHBERN, C\'ecile Engrand\INSTCSNSM, Marina Galand\INSTICL, Carsten G\"uttler\INSTMPS, Pierre Henri\INSTLPC2E\AND\INSTOCA, Alain Herique\INSTGRENOBLE, Stavro Ivanovski\INSTRIESTE, Rosita Kokotanekova\INSTESO, Anny-Chantal Levasseur-Regourd\INSTLATMOS, Kelly E. Miller\INSTSWRI, Alessandra Rotundi\INSTIAPS, Maria Sch\"onb\"achler\INSTIGPCH, Colin Snodgrass\INSTROE, Nicolas Thomas\INSTUNIBERN, Cecilia Tubiana\INSTMPS, Stefan Ulamec\INSTDLR, Jean-Baptiste Vincent\INSTDLRBERLIN

\vspace{+3cm}

\noindent
{\bf {\Large Affiliations}}
 \newline
\INSTIAPS Istituto di Astrofisica e Planetologia Spaziali, Istituto Nazionale di Astrofisica, via del Fosso del Cavaliere 100, 00133 Rome, Italy \\
\INSTUNIBERN Physikalisches Institut der Universit\"at Bern, Sidlerstr. 5, 3012 Bern, Switzerland\\
\INSTGRENOBLE Univ. Grenoble Alpes, Institut de Plan\'{e}tologie et d'Astrophysique de Grenoble, 38000 Grenoble, France\\
\INSTDAN Centre for Star and Planet Formation and Natural History Museum of Denmark, University of Copenhagen, DK-1350 Copenhagen, Denmark \\
\INSTTUBS Institut f\"ur Geophysik und extraterrestrische Physik, Technische Universit\"at Braunschweig, Mendelssohnstr. 3, 38106 Braunschweig, Germany \\
\INSTINAFFI INAF, Osservatorio Astrofisico di Arcetri, Largo E. Fermi 5, 50125, Firenze, Italy \\
\INSTJPL Jet Propulsion Laboratory/California Institute of Technology, 4800 Oak Grove Dr., Pasadena, California, 91109, USA \\
\INSTLISA Laboratoire Interuniversitaire des Syst\`{e}mes Atmosph\'{e}riques (LISA), UMR CNRS 7583,  Universit\'{e} Paris Est Cr\'{e}teil, Universit\'{e} de Paris, Institut Pierre Simon Laplace, 94000 Cr\'{e}teil, France\\
\INSTCSHBERN Center for Space and Habitability (CSH), Universit\"at Bern
Gesellschaftsstrasse 6, 3012 Bern, Switzerland\\
\INSTCSNSM Centre de Sciences Nucl\'{e}aires et de Sciences de la Mati\`{e}re (CSNSM) CNRS-IN2P3/Univ. Paris Sud, Universit\'{e} Paris-Saclay, B\^at. 104, 91405 Orsay Campus, France\\
\INSTICL Department of Physics, Imperial College London, Prince Consort Road, London SW7 2AZ, UK\\
\INSTMPS Max Planck Institute for Solar System Research, Justus-von-Liebig-Weg 3, 37077 G{\"o}ttingen, Germany \\
\INSTLPC2E LPC2E, CNRS, Universit\'e d'Orl\'eans, 3 Avenue de la Recherche Scientifique, 45071 Orl\'eans, France\\
\INSTOCA Laboratoire Lagrange, Observatoire de la C\^ote d'Azur, CNRS, UCA, 96 Boulevard de l'Observatoire, 06300 Nice, France\\
\INSTRIESTE INAF, Osservatorio Astronomico di Trieste, Via G.B. Tiepolo, 11 I-34143 Trieste, Italy \\
\INSTESO  European Southern Observatory, Karl-Schwarzschild Str. 2, 85748 Garching bei M\"unchen, Germany \\
\INSTLATMOS Sorbonne Universit\'e, LATMOS-IPSL-CNRS, BC 102, 4 place Jussieu, 75005 Paris, France\\
\INSTSWRI Southwest Research Institute, 6220 Culebra Rd, San Antonio, TX 78238, USA\\
\INSTIGPCH Department of Earth Sciences, ETH Zurich, Clausiusstrasse 25, 8092 Zürich, Switzerland\\
\INSTROE Institute for Astronomy, University of Edinburgh, Royal Observatory, Blackford Hill, Edinburgh EH9 3HJ, UK \\
\INSTDLR DLR, Space Operations and Astronaut Training, Linder H\"ohe, 51147 Cologne, Germany \\
\INSTDLRBERLIN Deutsches Zentrum f\"ur Luft- und Raumfahrt (DLR), Institut f\"ur Planetenforschung, Rutherfordstra{\ss}e 2, 12489 Berlin, Germany \\

\newpage

\noindent
\fcolorbox{ceruleanblue!90}{ceruleanblue!90}{\textcolor{white}{{\bf {\Large Executive Summary}}~~~'...Wise men know the comets come back' (V. Nabokov 1899-1977)}} 
\vspace{+0.1cm}

The {\it Giotto} mission was the first interplanetary probe ever flown by the European Space Agency (ESA). Selected in 1980 and flown in 1985, {\it Giotto} was the most daring of an international fleet of five spaceprobes, which triumphantly visited the comet of all comets, 1P/Halley, in March 1986. Earlier on, in 1984, ESA and NASA established a Comet Nucleus Sample Return Science Definition Team; their work led in 1993 to the selection of the {\it Rosetta} mission as the Planetary Cornerstone of ESA's long-term programme Horizon 2000. {\it Rosetta} was the first mission ever to land an automated laboratory ({\it Philae}) on the surface of a cometary nucleus and to accompany an errant body in its active phase through the Solar System. Launched in 2004, {\it Rosetta} ended its operational life on the 30th of September 2016.

{\it Giotto} and {\it Rosetta} have completely transformed our understanding of comets and have contributed hugely to give Europe a well-deserved leadership in the field of cometary studies. Selecting {\it Giotto} and {\it Rosetta}, Europe demonstrated the vision and the ambition to take the lead in the worldwide effort to learn about our origins. ESA has recently selected {\it Comet Interceptor}, a much smaller-scale mission (F-class) to encounter a yet-to-be-discovered Oort cloud comet, following a novel approach to expand our horizons on a limited budget. {\bf Europe has the opportunity to confirm and reinforce this leadership with an even more daring program: \textcolor{blue(pigment)}{\it AMBITION}, a mission to return the first-ever cryogenically-stored sample of a cometary nucleus to Earth.} 
The international context is very favourable as, after the recent selection by NASA of the next New Frontiers class candidate mission, no international agency (NASA, CNSA, JAXA) is presently planning a sample return mission from a cometary nucleus in the next decade.  NASA's New Frontiers candidate {\it CAESAR}, the unique direct competitor of \textcolor{blue(pigment)}{\it AMBITION}, was not selected and, even if {\it CAESAR} could still be a potential candidate for the next New Frontiers call, NASA's present timeline gives Europe a competitive edge with the \textcolor{blue(pigment)}{\it AMBITION} endeavour.

The numerous discoveries of the {\it Rosetta} mission are of great relevance not only for their scientific significance, but also because they provided essential insight into performing critical operations and measurements. This information provides better focus for the next set of fundamental questions that can be answered by the \textcolor{blue(pigment)}{\it AMBITION} project. 
{\bf This white paper, put together by a large scientific community ranging from cosmochemists to plasma physicists, presents compelling evidence that a mission capable of returning a cryogenic sample of a cometary nucleus to be analysed on Earth, and of studying the comet both remotely and in-situ, will be able to dig deeper into our past.}


\vspace{-0.1cm}
\textcolor{white}{\section{{\bf Introduction}}}

Comets and primitive asteroids are the leftover building blocks of giant planet cores and other planetary bodies, and fingerprints of Solar System's formation processes. They are composed of components that formed in the earliest stages of Solar System history. These objects have also preserved materials that predate the formation of the protoplanetary disk. Asteroids are mostly found in the inner Solar System, inside the orbit of Jupiter, where they are believed to have formed. Conversely, comets, which are stored in two main reservoirs in the outer Solar System, the Oort Cloud and the Kuiper Belt, are classically thought as ice-rich bodies formed beyond the snow line in the outer Solar System. The distinction between comets and asteroids has recently been blurred by the discovery of both ice-rich asteroids and active asteroids in the asteroid Main Belt \citep{2017A&ARv..25....5S}. From the analysis of cometary dust particles  using {\it Stardust} samples returned from comet 81P/Wild 2 or particles of 67P/Churyumov-Gerasimenko (thereafter, 67P) studied in situ by {\it Rosetta}, it is also now clear that comets and asteroids constitute part of a continuum in composition, pointing to an extensive transport of inner Solar System material into the outer regions in the early history of the Solar System \citep{2006Sci...314.1711B, {2016Natur.538...72F}}. The exploration of the different classes of minor Solar System bodies, including Main Belt comets and transition objects such as Centaurs, is of paramount importance to address Solar System formation and history, and the various processes which have altered these bodies since their formation.

Comets are of great scientific value because their ices sublimate when they approach the Sun, offering the opportunity to access the primitive volatile component of the solar nebula (the Solar System protoplanetary disk). Besides water, these ices are composed of both simple and complex molecules, and appear to be common to star-forming regions, suggesting a formation in the presolar cloud (the molecular cloud precursor of the Solar System; \citep{2011ARA&A..49..471M,2017MNRAS.469S.130A}). Unlike asteroids, comets do not seem to have been thermally or aqueously altered after accretion \citep{2015Sci...347a0628C,2016Icar..272...32Q}. In addition, comets have near-solar elemental abundances and large amount of organic matter found in the refractory material \citep{2017MNRAS.469S.712B}. {\bf Comets likely constitute the most primitive material in the Solar System}.

The space exploration of comets started in the 80's with the flybys of comet 1P/Halley. With its extensive payload, the {\it Giotto} spacecraft from ESA revealed properties of comets previously unimaginable, e.g., the incredibly dark surface of cometary nuclei and the large amount of organics in dust particles \citep{1986Natur.321..320K,1986Natur.321..336K}. NASA missions to comets 19P/Borrelly, 81P/Wild~2, 9P/Tempel~1, and 103P/Hartley~2 demonstrated an amazing diversity in the surface geology and yielded new insights into the bulk  properties of cometary nuclei, the nature of cometary dust, and the processes involved in cometary activity \citep{2002Sci...296.1087S,2005Sci...310..258A,2006Sci...314.1711B,2011Sci...332.1396A}. Analysis of the samples of comet 81P/Wild 2 collected and delivered to Earth by the {\it Stardust} spacecraft led to major findings, though the sample collection at 6.1 km/s left little chance for the survival of volatiles and organic matter \citep{2006Sci...314.1711B}. 

{\bf ESA's {\it Rosetta} mission}, the third cornerstone mission of the ESA programme Horizon 2000, {\bf is the most ambitious and sophisticated cometary space project attempted} \citep{2017RSPTA.37560262T,2017RSPTA.37560248B}. During its 2~years (Aug. 2014-Sept. 2016) escorting comet 67P, the 10 instruments aboard the spacecraft monitored and performed both in situ and remote analyses of the nucleus, atmosphere and solar-wind interaction continuously. {\it Rosetta} enhanced its science return by releasing {\it Philae} which landed on the nucleus surface: despite the non-nominal landing, the instruments aboard the lander were able to complete a subset of the planned measurements aimed at characterizing the comet nucleus's surface, subsurface and local environment \citep{2017RSPTA.37560248B}. {\bf {\it Rosetta} has provided many important and often unexpected results, as demonstrated by the number of publications} (presently $\sim$ 1000), with many in {\it Science} and {\it Nature} journals. A wealth of new knowledge was obtained on:

\begin{itemize}
\setlength{\parskip}{0pt} \setlength{\itemsep}{0pt plus 1pt}
\item the bulk and internal properties of the nucleus, with the demonstration that comets have very low density ($\sim$ 500 kg/m$^3$) and high porosity ($ > 70$ \%)  \citep{2015Sci...349b0639K,2019MNRAS.483.2337P}, and the evidence for layering in the uppermost parts of the nucleus which may be related to  primordial or post-collisional  accretion \citep{Massironi2015,2017A&A...597A..61J}; the nucleus interior is very homogeneous down to few-meter scales \citep{Herique2019};
\item the surface morphology, showing an unexpected diversity with large areas of the surface covered by airfall particles originating from the southern hemisphere \citep{2015Sci...347a0440T,2017MNRAS.469S.357K}. There are clear indications that no single process dominates the evolution of the surface morphology;
\item the distribution of water ice on the surface, and its diurnal and seasonal evolution (e.g., \citep{2015Natur.525..500D,2016Sci...354.1566F}).  The refractory-to-ice ratio in the nucleus exceeds 3, inconsistent with the paradigm that comets are very ice-rich (e.g.,  \citep{2019MNRAS.483.2337P,2019MNRAS.482.3326F}); 
\item the thermal and mechanical properties of the surface and subsurface layers; the very low bulk tensile strength is consistent with a primordial rubble-pile formed in the solar nebula \citep{2018A&A...611A..33A}; 
\item the physical properties of dust particles \citep{2018SSRv..214...64L}, with the discovery of both highly porous or even fractal-like and more compact aggregates made of $\sim$ 100 nm subunits consistent with  dust  growth starting by low-velocity    hierarchical accretion \citep{2015ApJ...802L..12F,2016MNRAS.462S.304M,2016MNRAS.462S.132F};
\item the solid organic matter ($\sim$ 50\% in mass) in dust particles which is bound in very large macromolecular compounds and is analogous to the insoluble organic matter in carbonaceous chondrites \citep{2016Natur.538...72F,2016MNRAS.462S.516H};
\item the volatile composition, with the detection of a wealth of simple and complex species, e.g.,  O$_2$, noble gases, heavy hydrocarbons, alcohols, and glycine \citep{2016SciA....2E0285A,2015Natur.526..678B}.  Isotopic measurements for H, C, O, Si, S, Ar, and Xe, and similarities and differences with primitive carbonaceous meteorites shed light on the presolar versus solar-nebula origin of cometary ices \citep{2018SSRv..214..106H}. Xenon isotopic abundances provide evidence that comets contributed to atmospheric Xe on Earth, or more generally to atmospheric noble gases \citep{2017Sci...356.1069M};
\item the importance of diurnal and seasonal illumination conditions on the different regions of the nucleus in driving the actual activity \citep{2016MNRAS.462S...2F};
\item the evolution of the interaction of a comet with its surrounding space environment, from the birth of an induced magnetosphere and a nearly collisionless cometary plasma, to a chemically-active, colder, unmagnetized cometary plasma near perihelion \citep{2018A&A...616A..51E,2017MNRAS.469S.268G,2018A&A...618A..77H,Nilsson2019}.     

\end{itemize}             
\vspace{-\topsep}

\vspace{+0.5cm}

\noindent
Although the {\it Rosetta} mission brought cometary science to a new level  
of maturity and improved our undestanding of the early Solar System, a number of fundamental questions that can only be achieved by space missions remain unresolved.

\vspace{-\topsep} 
\begin{itemize}
\setlength{\parskip}{0pt} \setlength{\itemsep}{0pt plus 1pt}
\item How did cometary materials get assembled? What are the building blocks of cometary nuclei? 
\item Which post-planetesimal evolutionary paths need to be considered? Are comets collisional fragments or primordial planetesimals?
\item Which cometary components pre-date the formation of the Solar System? 
What is the nature of the refractory organic materials? How and where did these components form?
\item Are there differences in physical/chemical properties in the comet populations? What are the commonalities and differences between comets and primitive asteroids?  Are these distinct populations? Do comets, Main Belt comets and  “active” asteroids have different primordial reservoirs?
\item How does comet activity work?  How do surface and coma observations reconnect with the pristine, deep interior?
\item How are the dusty coma, the surrounding plasma, and the nucleus interacting together? Do interactions with the solar wind influence the activity and evolution of comets?
\item What was the role of comets in the delivery of volatiles and prebiotic compounds to early Earth?
 \end{itemize}             
\vspace{-\topsep}

\vspace{+0.5cm}

\noindent
{\bf The holy grail of cometary spacecraft missions is the return to Earth of a cryogenic sample extracted from deep within the nucleus}. It is worth mentioning that the original {\it Rosetta} mission was a comet-nucleus sample return (CNSR) and that significant important studies were carried out towards these goals. CNSR missions allow the ultimate level of detailed study of extraterrestrial samples, although they are technically challenging and expensive. {\bf In situ studies with static or mobile landers, equipped with a sophisticated payload and drilling systems, remain a necessary complement to sample return because of inherent problems with maintaining sample integrity, in addition to enhance the science return from the sample by providing the full context}. We present the major scientific questions resolvable by future comet missions in Section 2, and propose  a few missions for the ESA Voyage 2050 long-term plan in Section 3. The international context of comet space and Earth-based exploration is presented in Section 4.


\textcolor{white}{\section{{\bf Top Level scientific questions}}}

\subsection{How and where did cometary materials get assembled? Which post-planetesimal evolution paths need to be considered?} 
\label{sec:assemblage}

\begin{figure}
\centerline{
\includegraphics[height=7cm]{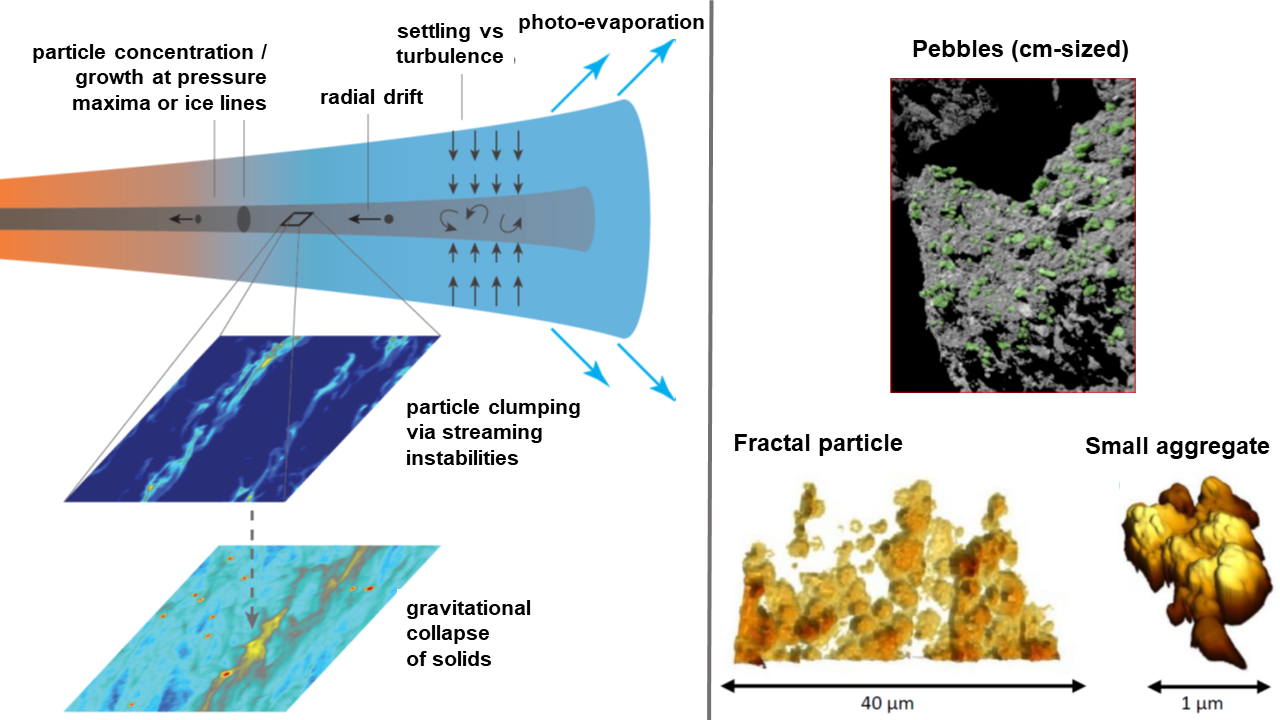}} 
\caption{{\it Left}: Processes in protoplanetary disks giving rise to gas streaming instabilities and the formation of pebbles and planetesimals \citep{2019SAAS...45....1A}; {\it Right}~(67P building blocks): pebbles on the surface imaged by the Comet Infrared and Visible Analyser (CIVA) onboard {\it Philae} (top) \citep{2016MNRAS.462S..23P} and 3D rendered images obtained with the Micro-Imaging Dust Analysis System (MIDAS) (bottom) \citep{Mannel2019}.}
\label{fig:disk}
\end{figure}

It is undisputed that the formation of planetary bodies in the solar nebula started with the coagulation of fine dust particles induced by low-velocity sticking collisions \citep{Weidenschilling1977}. Recent numerical models of the physical agglomeration process, taking into account the latest findings from laboratory experiments (e.g.,  \citep{Guettler2010}), show that mm- to cm-sized agglomerates (hereafter, termed “pebbles” in accordance with the literature) form within a few hundred to a few thousand orbital timescales \citep{Lorek2018,Zsom2010}. Due to the systematic increase in collision velocity with increasing agglomerate mass, larger agglomerates no longer simply stick after a collision. However, some of the collisions may still lead to a further gain in mass, even though the sticking threshold has been overcome, e.g., by mass transfer in high-velocity collisions (see \citep{Guettler2010} and references therein). \citet{Davidsson2016} described how cometary nuclei might have formed in such a scenario and compared it to the findings of {\it Rosetta}. An alternative scenario was presented by \citet{Blum2017} who collected evidence from {\it Rosetta} data that the comet nucleus might still consist of the primordial pebbles, with fractal dust aggregates captured between the pebbles \citep{Fulle2017}, which indicates that it formed by the gentle gravitational collapse of a concentrated cloud of pebbles \citep{Johansen2007} (Fig.~\ref{fig:disk}).

Without further collisional evolution, these two formation scenarios predict very different physical properties of the resulting planetesimals, e.g., porosity and tensile strength \citep{Blum2018}. Morbidelli and Rickman \citep{Morbidelli2015} argued that the planetesimals in the outer Solar System underwent one or more \mbox{(sub-)catastrophic} collisions during the past 4.5 billion years. In this case, the primordial planetesimal properties might have been overwritten by the fragmentation and re-accretion events. However, major parts of the reassembled cometary nucleus might still bear the initial morphology so that not necessarily all information about the underlying formation mechanism might have been lost \citep{Schwartz2018}. Additional difficulties arise when trying to identify the properties and structures that may have evolved via other forms of secondary processing (including, but not limited to, distortion or fragmentation caused by spin-up and/or planetary tides; solar and/or radiogenic heating and associated vapor diffusion, recondensation and irreversible phase transitions; gravitational compaction). Indeed, {\it Rosetta} showed a wide variety of different structures and textures on 67P nucleus surface \citep{2015Sci...347a0440T}  and complex surface changes due to activity \citep{2017Sci...355.1392E,2017MNRAS.469S..93F}. 
 It is thus  mandatory to isolate features indicative of the initial growth process of the planetesimal precursors of cometary nuclei. For example, the {\it Deep Impact} and {\it Rosetta} missions demonstrated that stratigraphic layering is an important structural component in some comet nuclei \citep{Belton2007,Massironi2015}, but it is unknown whether layering forms during the growth of the body or at later stages.

The complex interplay between formation and evolutionary processes requires both advanced modeling and dedicated laboratory work to arrive at realistic predictions on the physical and structural properties of bodies formed in different ways. Processes such as the transition from fractal to non-fractal (but porous) aggregate growth domains, the dependence of growth barriers on physical and structural dust particle properties, the interplay between erosional and growth processes in moderate-velocity impacts, or advanced modeling of collisions among porous and hierarchically structured bodies must be the focus of future studies. An anticipated comet mission with the aim of revealing the formation and evolution processes of planetesimals can only be successful if all the above-mentioned aspects are properly taken into account. Only a holistic approach to revealing the structural properties of the entire comet nucleus (or at least considerable parts of it) can lead to the desired results. {\bf The ability to investigate the interior of the nucleus down to a depth of several tens to hundreds of meters with a resolution in the mm-cm range may be the essence of this investigation}. It is also important to understand whether structures seen on such size scales are representative for the body as a whole, or if regional variability is substantial. {\bf In this respect, a successful mission will need to measure the morphological and compositional  properties (particle  sizes, dust/ice/organic mixing, arrangement on different scales) of a sample that never experienced modification by cometary activity. This implies that material from the deep interior needs to be studied on a wider range of size scales and with a composition-probing capability.} A careful assessment of the relative merit of remote-sensing instruments (e.g., ground-penetrating radar), as well as in situ investigations (drill cores, mole probes, penetrators, impact excavation) is required. The possibility of exploiting the exposure of nucleus interiors following comet fragmentation or splitting events, as well as large-scale outbursts, should be considered.

\subsection{What is the presolar heritage of cometary materials, versus a solar nebula  origin?}
\label{sec:heritage}

Through their mineralogical, chemical, and isotopic composition, comet nuclei document 
environmental conditions and processes occurring from the protostellar collapse phase to the protoplanetary disk phase. The protostellar collapse phase includes the prestellar core formation from the parent molecular cloud, its collapse, and the bulk solar accretion phase as the protosun evolved through the protostellar Class 0 and Class I stages \citep{1977ApJ...214..488S,1994ApJ...420..837A}. Physical processes such as gas compression and associated heating (e.g., in shocks), inward accretion flows, disk winds, and bipolar flows, take place \citep{1996ARA&A..34..111B,2006ApJ...641..949B,2011MNRAS.413.2767M}. Icy dust particles feeding the disk are affected by thermal evaporation, photodesorption, irradiation, and sputtering in strong shocks. Complex organic molecules have been observed as early on as in the prestellar cores with their formation stemming from grain-surface chemistry (e.g.,  \citep{2012A&A...541L..12B}. Physico-chemical models suggest that their formation is further enhanced during protostellar collapse and continues within the protoplanetary disk midplane \citep{2014A&A...563A..33W,2016MNRAS.462..977D}. To what extent does molecular complexity develop in the protoplanetary disk phase versus being inherited from prestellar ices? How do nebular processes, such as stellar UV, X-ray and cosmic rays, chemically alter volatiles and dust particles? The chemical inventory of the dust particles entering the protoplanetary disk depends on several parameters: how long did the prestellar core spend in the cold-dense phase, what was the temperature in the prestellar core, and what are the characteristic fluences of  FUV and cosmic rays? What was the chemical composition of the prestellar core of the Solar System?

\begin{figure}
\vspace{-0.5cm}
\parbox[t]{.5\columnwidth}{
\includegraphics[angle=0,width=12.5cm]{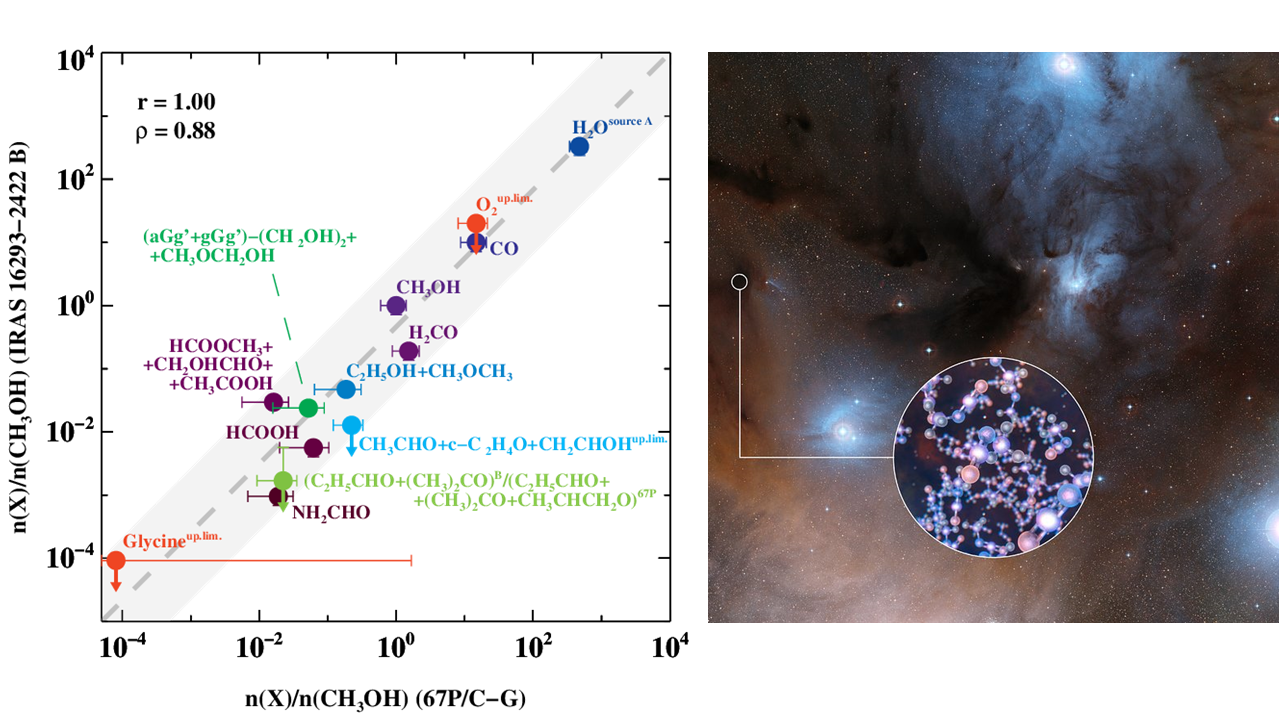}}\hfill
\parbox[b]{.25\linewidth}{\vspace{-1.5cm}\caption{{\small {\it Left}: Abundance of CHO bearing molecules relative to methanol in IRAS 16293-2422 B versus that measured in comet 67P (arrows indicate upper limits) \citep{Drozdovskaya2019}. {\it Right}: the Rho-Ophiuchi star-forming region harbouring IRAS 16293-2422 (ESO/Digitized Sky Survey 2/L. Cal\c cada). \label{classIvscomets}}}}
\end{figure}

Over the last few years, thanks to large programs on astrochemistry (using, e.g., the Atacama Large Millimeter Array, ALMA), a chemical census of the gas-phase has been obtained (and modelled) along the disk evolution from the protostellar phase to planet-forming disk. At the same time, the Rosetta Spectrometer for Ion and Neutral Analysis (ROSINA) onboard {\it Rosetta} identified many new cometary volatiles in comet 67P, including both complex organics (e.g., alcohols, long-chain hydrocarbons, glycine \citep{2016SciA....2E0285A}) and unexpected simple species such as O$_2$ \citep{2018A&A...618A..11T}. A comparative study of 67P with the closest Solar-like system that is still in its infant embedded phase of formation, namely the Class 0 low-mass protostar IRAS 16293-2422 B, shows striking correlations for CHO-, N- and S-bearing species suggesting {\bf inheritance from the presolar phase} \citep{Drozdovskaya2019} (Fig. \ref{classIvscomets}). A similar correlation is observed with the Class I hot-corino SVS13-A ($\sim$ 10$^5$ yr old) for complex species such as  NH$_2$CHO and HCOOCH$_3$ \citep{2019MNRAS.483.1850B}.  The higher O$_2$ content in 67P with respect to methanol in comparison to IRAS 16293-2422 B suggests a higher temperature in the prestellar core of the Solar System, with respect to the core from which IRAS 16293-2422 B formed \citep{2018A&A...618A..11T}. These findings call for further investigations. Especially important for a future mission will be the measurement of {\bf isotopic ratios, as key diagnostics of formation conditions}. These have been obtained in 67P and other comets with significant accuracy only for a small number of species, e.g., D/H in H$_2$O, HCN and H$_2$S, $^{18}$O/$^{16}$O in H$_2$O and CO$_2$ \citep{2015SSRv..197...47B,2017A&A...605A..50H,Schroeder2019},  $^{33}$S/$^{32}$S and $^{34}$S/$^{32}$S in H$_2$S, CS$_2$ and OCS  \citep{2015Sci...347A.387A,2017MNRAS.469S.787C,2017RSPTA.37560253A,2018SSRv..214..106H}. Whereas the high HDO/H$_2$O and D$_2$O/HDO ratios in 67P suggest a presolar heritage, this has to be confirmed in the light of recent findings on the D/H diversity in comets \citep{2019A&A...625L...5L}.

\begin{figure}
\centerline{\includegraphics[height=4cm]{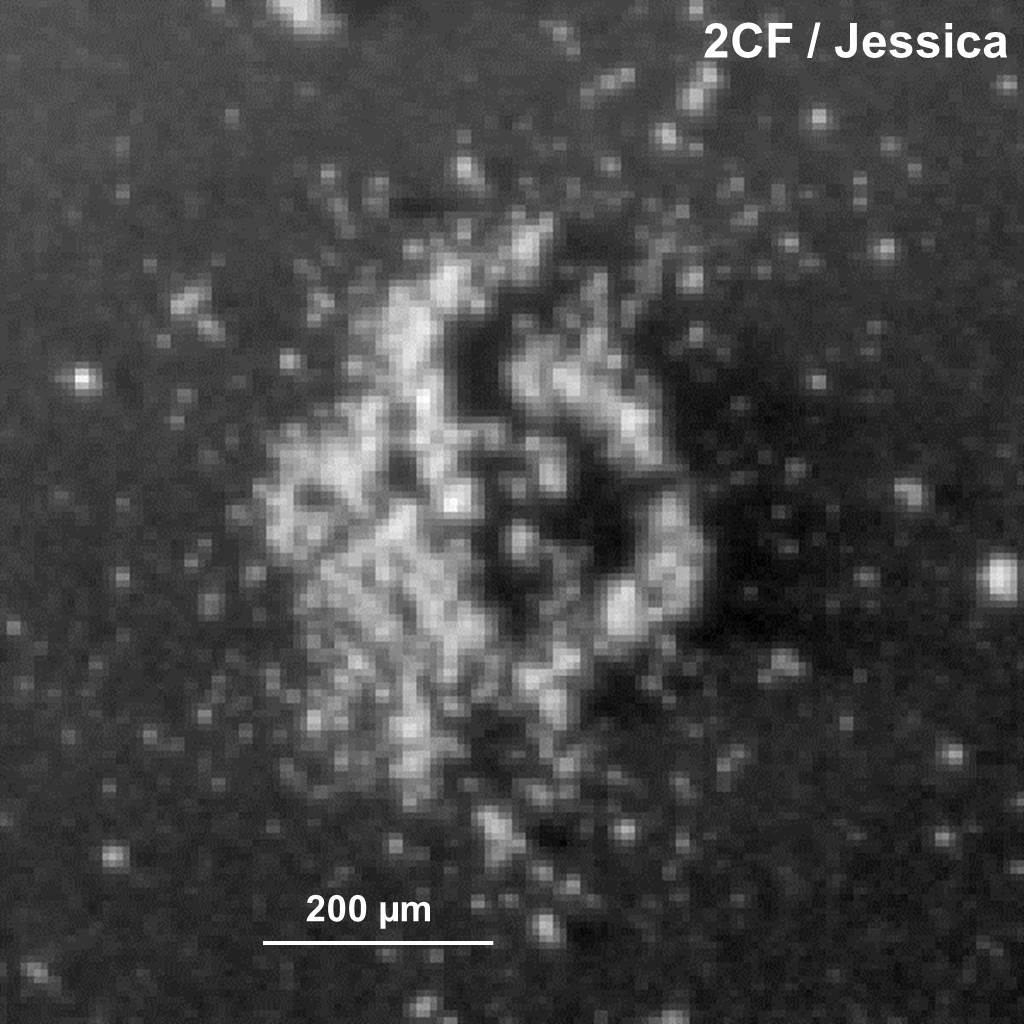}~\includegraphics[height=4cm]{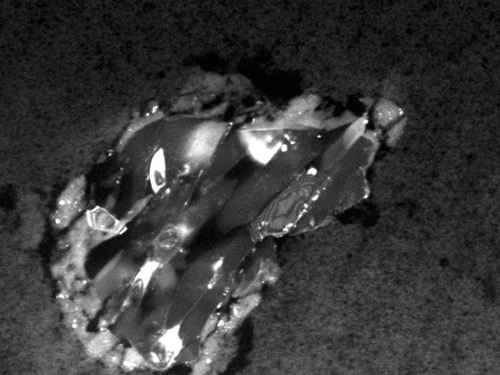}~\includegraphics[height=4cm]{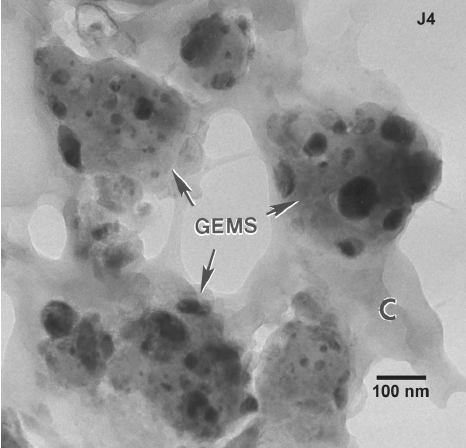}~\includegraphics[height=4cm]{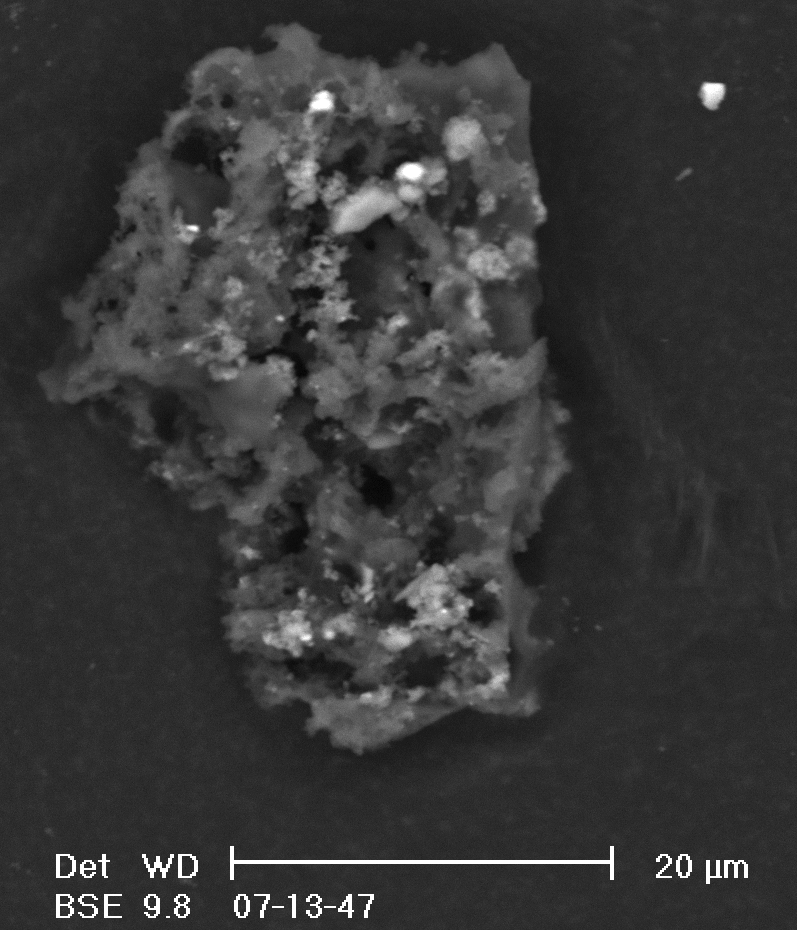}}
\caption{From left to right: 1) dust particle of 67P collected by COSIMA/Rosetta (credit: ESA/Rosetta/MPS); 2) 81P 2-$\mu$m particle  collected by {\it Stardust} made of Mg-rich crystalline olivine (credit JPL/NASA); 3) transmission electron micrograph of GEMS (Glass with Embedded Metals and Sulfides) embedded in their carbonaceous matrix from a potential cometary particle collected on Earth (CP-IDP) \citep{1994Sci...265..925B}; these glassy inclusions could have a presolar origin; 4) backscattered electron micrograph of an Ultracarbonaceous Antarctic MicroMeteorite (UCAMM) from the Concordia Collection (CSNSM, France); these particles are dominated by organic matter (dark phase) with minor mineral components (small and light phases). UCAMMs most probably have a cometary origin. }\label{fig:sample}
\end{figure}

{\bf Cometary dust is rich in refractory organic matter} (50\% in mass) \citep{2017MNRAS.469S.712B}. The identification and elemental characterization of CHON-rich dust particles were first performed in the coma of 1P/Halley \citep{1988Natur.332..691J,1992Natur.359..810L}. The {\it Rosetta} mission, and in particular instruments such as mass and IR spectrometers (ROSINA, Cometary Secondary Ion Mass Analyser (COSIMA), Visible and Infrared Thermal Imaging Spectrometer (VIRTIS)), allowed further characterization of this organic matter. Cometary organics  share some similarities with chondritic insoluble organic matter, although  they are not identical (e.g., \citep{2016Natur.538...72F,2016Icar..272...32Q}). 
Cosmic dust particles collected on Earth (mainly in the stratosphere and in the polar regions) contain material of probable cometary origin in the form of `chondritic porous interplanetary dust particles' (CP-IDPs, e.g., \citep{2008Sci...319..447I})  and `ultracarbonaceous Antarctic micrometeorites' (UCAMMs, e.g., \citep{2010Sci...328..742D}) (Fig.~\ref{fig:sample}). CP-IDPs and UCAMMs also contain $>$ 50\% in mass of organics.


Further chemical and isotopic characterization of cometary organics in the laboratory is necessary to establish a deeper comparison, and to decipher the origin(s) of cometary and chondritic organics, which are highly debated in the community. Different scenarios are advocated for chondritic organics: interstellar heritage, formation in the solar nebula, and parent body formation (e.g., \citep{2017ChEG...77..227A}). Isotopic fractionation measured in comets and chondrites are lower than those found in some species in interstellar clouds. This shows that primitive organics in the Solar System are either a mixture of interstellar matter and materials processed in situ, or produced in the outer regions of the solar nebula (e.g.,  \citep{1999ApJ...526..314A}). It has been suggested that the large organic molecules responsible for the Diffuse Interstellar Bands (DIBs) could be preserved in comets \citep{2017MNRAS.469S.646B}. Determining the interstellar vs. solar nebula origin of cometary organic matter requires  knowledge of the exact H, C, and N isotopic compositions of cometary organics, as well as the understanding of the evolution of organics from dense molecular clouds to protoplanetary disks. {\bf The direct comparison between cometary and chondritic organics (e.g., elemental and isotopic compositions, functional chemistry, structure, and texture) would also allow to address the question of either a unique or multiple reservoir(s) of organics in the Solar System } (at least 2 types of organic matter have been identified in UCAMMs \citep{2018LPI....49.2015E}). This comparison would, in turn, enable better understanding of the asteroid-comet continuum and potentially of the dynamics of a protoplanetary disk by constraining exchanges between the inner and outer Solar System. This latter point could also be addressed through the search for and characterization of high-temperature components, such as chondrules and CAIs, in cometary samples (Sect. 2.3). 

For comet 67P analyzed by {\it Rosetta}, glycine was identified in the coma by ROSINA \citep{2016SciA....2E0285A}, whereas low-molecular weight organic compounds were not found in the COSIMA spectra \citep{2016Natur.538...72F}. The absence of identification by COSIMA of organics comparable to the so-called soluble organic matter (SOM) in chondrites might indicate the presence of such soluble organics at a lower abundance in comparison to that of chondrites. Although its origin is debated (interstellar, solar, parent body origin), the SOM  is more abundant in the chondrites having experienced more intense aqueous alteration (e.g., \citep{2006mess.book..625P}). Comets have escaped such secondary processes, as suggested by the lack of evidence for hydrated minerals in cometary dust particles \citep{2016Icar..272...32Q}, although they were experienced by asteroids. {\bf The characterization of SOM in comets will give new constraints on the origin of chondritic soluble organics.} 

Although comets are rich in volatiles and solid organics, they are depleted in nitrogen in comparison to the solar nebula (e.g., \citep{2000Icar..146..583C}). However, the recent identification of ammonium salts at the surface and in the dust of comet 67P \citep{Poch2019,Altwegg2019} sheds new light on this issue: these salts  could provide the heretofore-missing link between comets and the parent interstellar cloud. While this potential new nitrogen reservoir was proposed from {\it Rosetta} data, no precise characterization of this reservoir is currently possible (such as the relative abundance of the different salts, and their isotopic composition). This would necessarily require the sampling of the cometary surface and its study in laboratories.

{\bf Presolar grains in primitive extraterrestrial materials} are identified by large isotopic anomalies compared to the solar composition (e.g., \citep{2004ARA&A..42...39C}). {\bf The nature of presolar grains gives access to the specific types of stellar sources (e.g., supernovae, red giant stars) in whose mass  outflows  the  grains  condensed}, and their residence time in the interstellar medium.  They  also  yield  insights into stellar nucleosynthesis at an unprecedented level of detail \citep{2004ARA&A..42...39C}. Their relative abundance in meteorites and comets gives insights into processes potentially acting both in the inner and outer regions of the protoplanetary disk. Presolar grains are found at a level of ~500 ppm in meteorites \citep{Zinner2005}, and up to ~1.5\% in IDPs collected during the Grigg-Skjellerup dust stream Earth's encounter in 2003 \citep{2009E&PSL.288...44B}. In comet 81P/Wild 2 samples returned by the {\it Stardust} mission, the abundance of presolar grains, after correction for destruction upon impact in the aerogel at 6.1 km/s, is ~650 to 900 ppm, being largely dominated ($>$90\%) by O-rich grains \citep{2013ApJ...763..140F}. This low level (rather comparable to that of meteorites) was found in contrast with measurements of silicon isotopes from comet 67P dust particles by ROSINA, which show an overall large depletion of heavy Si isotopes in the dust, suggesting a large fraction of presolar Si-bearing minerals \citep{2017A&A...601A.123R}.

Further characterisation of the composition of cometary dust, both for the organic and  mineral part, is clearly needed. Remote sensing and in situ analyses suffer from a lack of measurement context, and are not as precise and accurate as laboratory analyses. {\it Stardust} samples returned from comet 81P/Wild 2 suffered severely  from their hypervelocity collection, hampering meaningful understanding of the composition and structure of the matrix and organic matter, because only refractory components were retrieved. CP-IDPs and UCAMMs may have suffered from heating during atmospheric entry, although their porous structure could have limited their alteration. {\bf A cometary sample return would give the only opportunity to retrieve the fragile and low temperature components of cometary matter that are so far unknown in collections of primitive extraterrestrial matter, and to investigate a representative amount of material, giving access to the least abundant species (soluble organics, noble gases). A cryogenic sample return would also allow preservation of semi-volatiles, e.g., salts. }

\subsection{What do comets tell us about large-scale mixing and dynamical processes in the early Solar System?} 
\label{sec:mixing} 

During the protostellar collapse phase, dust particles located near the protostar experience harsh conditions. Amorphous silicates from the pre-stellar core anneal and crystallize, some solids evaporate and recondense entirely, and evolve into the granular mixtures of Calcium-Aluminum-Rich Inclusions (CAIs), agglomeratic olivines, amoeboid olivine aggregates, chondrules, and matrix material that are well-known from chondritic meteorites \citep{2004ChEG...64..185K,Jones2005,Mac2005,2012GeCoA..79...79R}. Infrared cometary spectra and analyses of {\it Stardust} samples of comet 81P/Wild 2 showed that such particles together with metals and sulfides were transported  to large heliocentric distances \citep{2000Icar..143..126W,2006Sci...314.1711B}, potentially accreting processed organics and ices on their way. Once there, they mixed with relatively unprocessed amorphous presolar minerals, organics, and ices (Sect. 2.2).
Indeed, the mineral components of comet 81P/Wild 2 samples share many similarities with that of primitive carbonaceous chondrites that are thought to originate from dark asteroids, including CAIs and chondrules \citep{2014AREPS..42..179B}. Cometary dust particles and cosmic dust particles of probable cometary origin that are collected on Earth (CP-IDPs and UCAMMs) contain both crystalline silicates and amorphous silicates of possible presolar origin \citep{2000Icar..143..126W} (Fig.~\ref{fig:sample}). The presence of crystalline and high-temperature minerals in cometary matter, sharing similarities with their counterparts in meteorites, has led to the notion of {\bf comet - asteroid continuum} \citep{Gounelle2011}. Cometary dust, on the other hand, contains much more carbon than carbonaceous meteorites (50\% carbon in mass, and even more in CP-IDPs and UCAMMs,  versus 5\% at most in meteorites) \citep{2017MNRAS.469S.712B}, in a form that is close to (but not identical to) insoluble organic matter in meteorites \citep{2017EGUGA..1912953F,Isnard2011,2017MNRAS.469S.506F,2018A&A...609A..65D}.

Multiple  mechanisms  have  been  proposed  to explain the  outward transport of solids in  the  solar  nebula prior to being accreted into planetesimals. Two leading theories are: i) solids were launched above the disk by winds that are able to decouple them from the gas   flow,  and  rained  back  down  onto  the  nebula  at  different  radial  locations \citep{1996Sci...271.1545S, 2009M&PS...44.1663C}; ii) processes operating within the nebular gas (turbulence, photophoresis, gravitational torques)  allowed solids to migrate outwards through the solar nebula \citep
{2002A&A...384.1107B,2008E&PSL.268..102B,2007A&A...466L...9M} (Fig.~\ref{fig:disk}). Alternatively, high-temperature materials formed on the disk surface far from the star during episodic accretion bursts \citep{2009Natur.459..224A}.  Depending on the transport mechanism, dust particles were exposed to very different environments during their transfer to the outer disk. Timescales for incorporation into planetesimals also differ. Determining which of the proposed mechanisms was dominant requires information on the degree of silicate crystallinity, the relative content of high-temperature material, their variation among different comet populations, and among primitive Solar System bodies formed at different places in the solar nebula.  Also important for constraining transport mechanisms is determining {\bf how and at what scales low- and high-temperature materials are hererogeneously or homogeously  mixed together.}

The search for extinct $^{26}$Al in cometary samples could allow (in the chronological interpretation) a precise dating of the crystallization time of cometary minerals. The absence of $^{26}$Al excess in {\it Stardust} samples measured so far could be interpreted as a late formation of minerals in comet 81P/Wild 2 dust \citep{2010Sci...328..483M,2012ApJ...745L..19O,2015E&PSL.410...54N}. Other dating systems (U-Pb, Rb-Sr, Nd-Sm,...) could provide information about the formation ages and timing of possible later processes occurring on the comet. {\bf Establishing a precise chronology in the early Solar System is only possible through analysis of pristine samples in the laboratory.}

Comets have historically been distinguished from asteroids by the extended coma and tail that they develop when approaching the Sun. This distinction was blurred in 2006 by the discovery of a significant population of asteroids with comet-like activity in the outer Main Belt (the so-called {\bf Main Belt Comets} (MBCs) \citep{2006Sci...312..561H}). The most recent estimate gives a population of 140--150 active bodies, more than could have found their way from known comet reservoirs by complex gravitational interactions \citep{2015Icar..248..289H}. In addition, ice signatures have been identified on the surface of several asteroids (e.g., \citep{2010Natur.464.1320C,2019Icar..318...22C}), and water vapor was detected around the ice-rich asteroid Ceres \citep{2014Natur.505..525K}. The implied presence of water in the Main Belt within the snow line in the protoplanetary disk indicates that significant migration of the planetesimals has taken place early in the Solar System.  This is in line with dynamical models aimed at understanding the formation of the terrestrial planets and the present architecture of the Solar System. Because of interactions with gas and/or planetesimals, giant planets underwent episodes of migration, scattering bodies by many astronomical units from the location of their formation. The Grand Tack model, describing the inward then backward migration of Jupiter and Saturn in the gaseous nebula, not only helps in explaining the properties of the terrestrial planets, but also the structure of the asteroid belt and the presence of S- and C-type asteroids associated with ordinary and primitive carbonaceous chondrites, respectively \citep{2011Natur.475..206W,2014Natur.505..629D}. The formation of the two main comet reservoirs, the Oort Cloud and the Scattered Disk,  sources of the long-period and short-period dynamical families, respectively, is believed to 
have resulted from the gravitational interaction of the three outermost external planets with a massive disk of planetesimals extending out to 30 AU, which made them move outwards, dispersing the disk (so-called Nice model \citep{2011Natur.475..206W}). This means that the diverse comet and asteroid reservoirs probably include bodies formed at different locations in the Solar System. {\bf The suggestion that volatile-rich carbonaceous chondrites originate from the outer Solar System or even from comets is an interesting hypothesis that can be tested by a comet nucleus sample return} \citep{2008ssbn.book..525G}.  

{\bf The characterization of asteroid-comet continuum objects, like MBCs, is of prime importance to test Solar System evolution scenarios and for constraining the distribution of water in the early Solar System. Because these objects are weakly active and small, their investigation requires space exploration. The detection of water vapor 
and its sources would attest to the presence of buried ice, and constrain the activity-triggering mechanism. The measurement of the D/H ratio in water is a prime objective for constraining their formation region, together with physical and chemical properties that can be compared to asteroids and comets properties.}

\subsection{How does comet activity work? How do surface and coma observations reconnect with the pristine, deep interior?} 
\label{sec:activity}

Comets show several types of activity depending on the nature of the gas and dust involved, flux and distribution of the sources (localized or diffuse) that are related to the thermophysical and chemical properties of the comet nucleus. {\it Rosetta} and previous missions greatly improved our understanding of how comets work, but there are still open questions.  Cometary activity can originate from the immediate surface and from the subsurface, and the associated phenomena are different. Also, the mechanisms driving the gas and dust activity can be different and can operate at different times and heliocentric distances.

\begin{figure}
\centerline{\includegraphics[height=6cm]{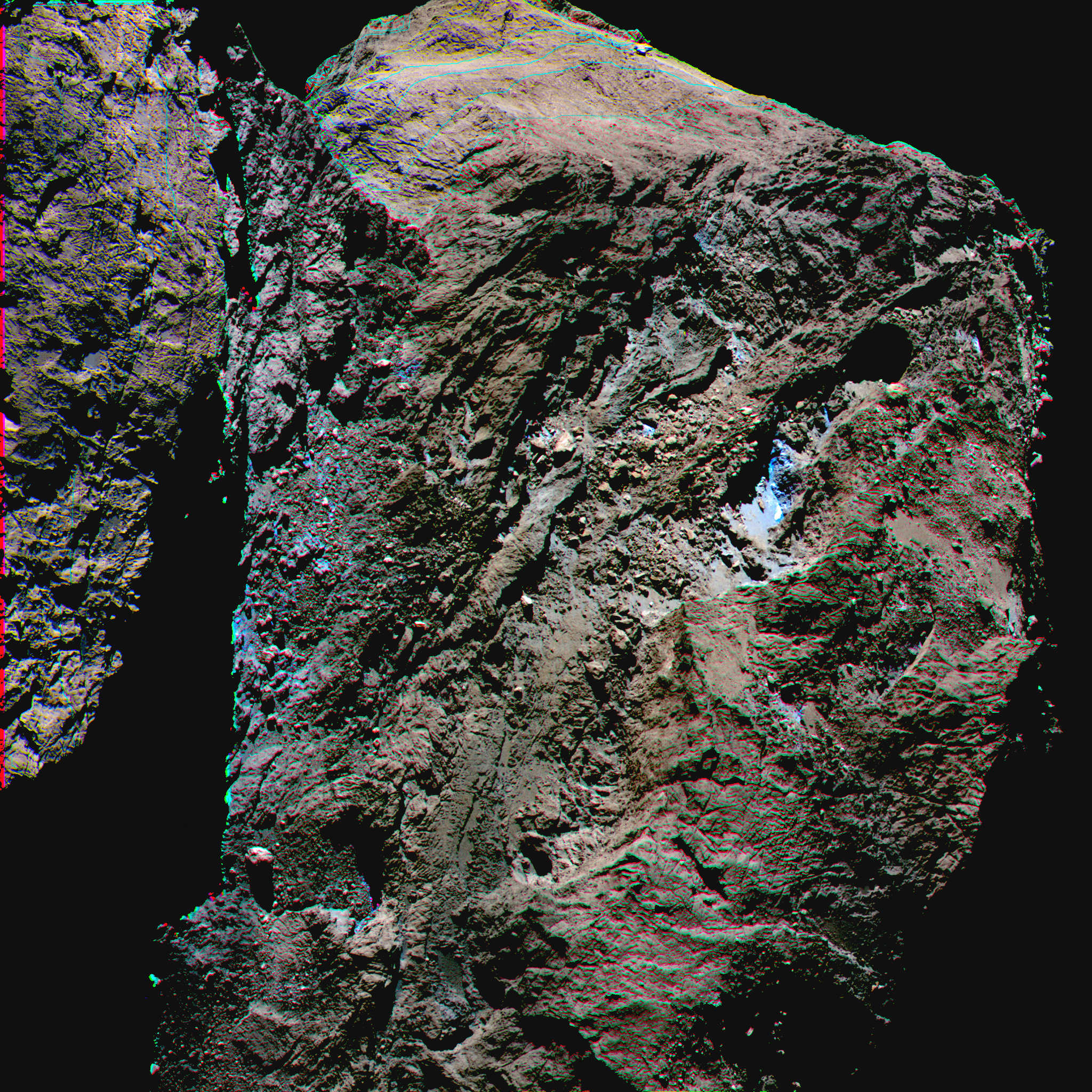}~\includegraphics[height=6cm]{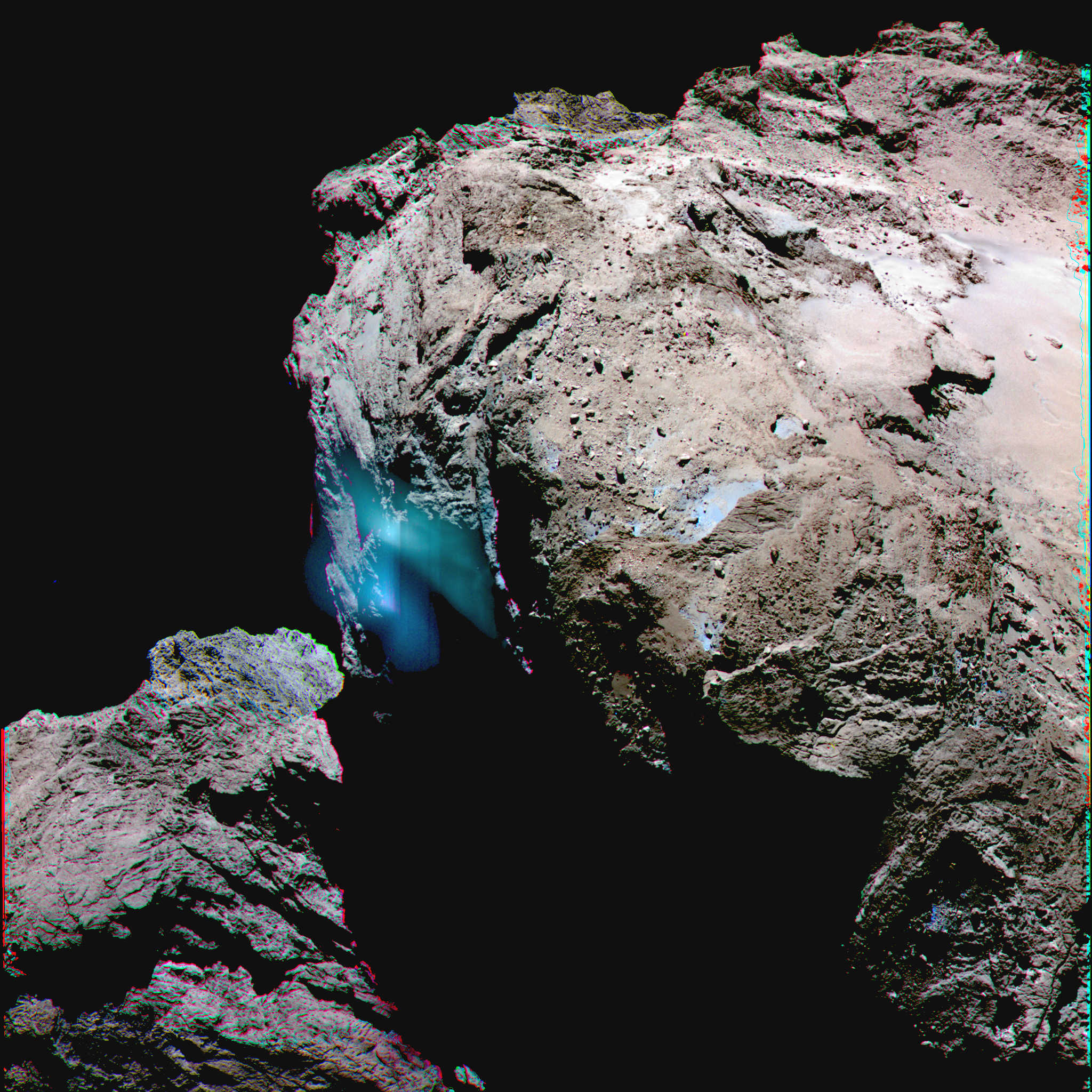} \includegraphics[height=6cm]{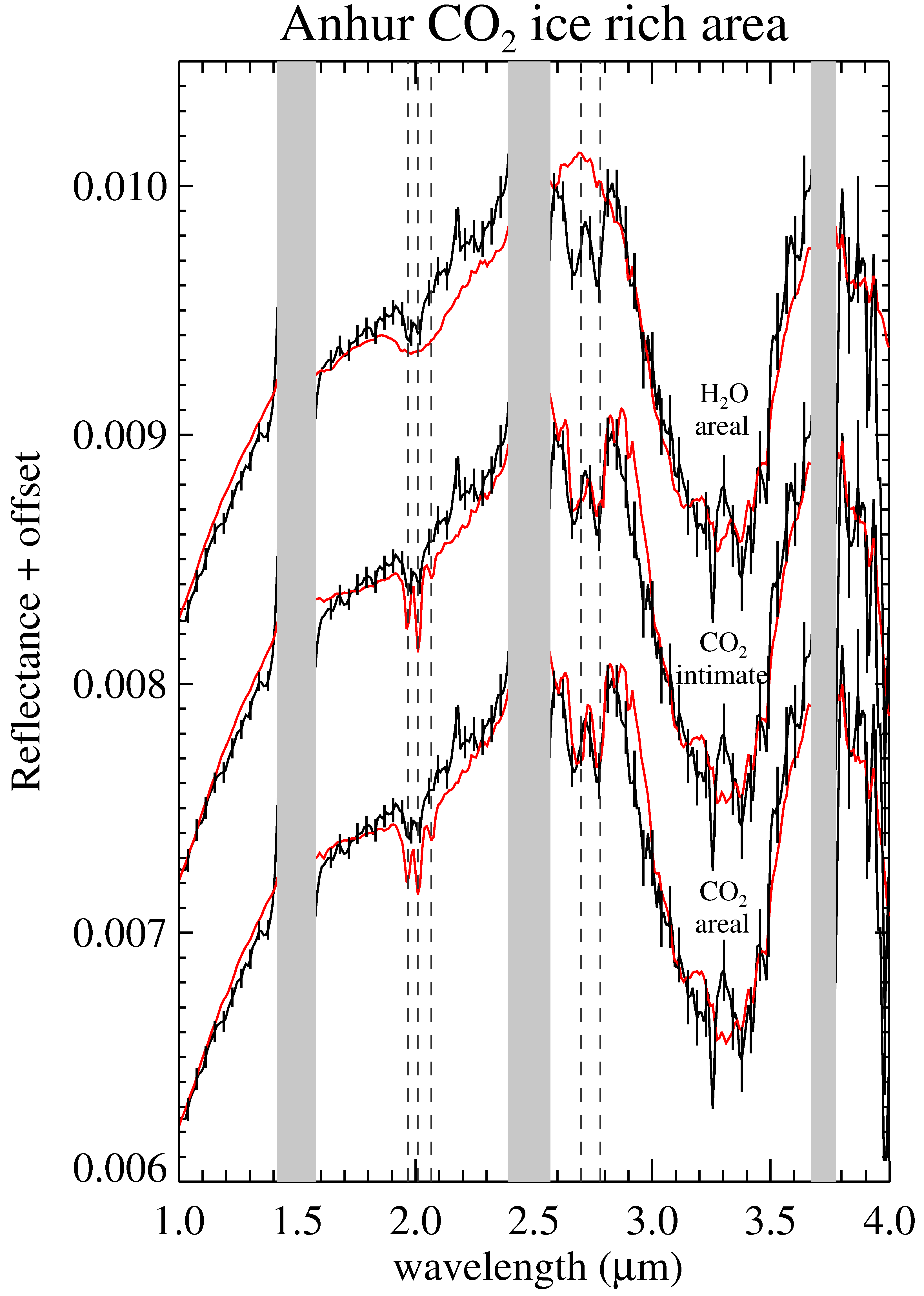}}
\caption{{\it Left and center}: RGB maps of 67P's Anhur region obtained with the Optical, Spectrocopic and Infrared Remote Imaging System (OSIRIS), showing exposed water ice (blue patches) on the dark surface, and jet activity \citep{2017MNRAS.469S..93F}. {\it Right}: VIRTIS spectrum showing the transient signature of surface CO$_2$ ice on Anhur (dashed vertical lines) \citep{2016Sci...354.1563F}. A combination of water ice,  organics and salts is responsible for the broad 3.2 $\mu$m band \citep{2015Sci...347a0628C,2016Icar..272...32Q,Poch2019}.   }
\label{fig:activity}
\end{figure}

Different types of local activity have been observed, including i) diffuse activity from the nucleus and ii) localized activity (jets, outbursts; Fig.~\ref{fig:activity}). Jets seem to be related to surface topography (e.g., sinkholes or  fractures), but also to sudden changes in thermophysical condition of the surface, such as  shadow boundaries (e.g., the terminator line, shadows from topographic reliefs) 
\citep{2015Natur.525..500D,2016MNRAS.462S.184V,2016A&A...587A..14V,2016ApJ...823L..11K,2016A&A...586A...7S}). Diffuse activity is more difficult to trace, because it is not clearly linked to surface features, but is more related to the overall nucleus thermochemical state.

Several mechanisms have been recognized to play a major role in the activity of comets. On a global scale, solar input and seasonal and diurnal variations act on the nucleus, driving the gas activity, which starts with the sublimation of the most volatile ices at large heliocentric distance, increasing close to perihelion with the release of the less volatile species \citep{2015Sci...347a0276H,2016A&A...589A..45M,2016MNRAS.462S.491H,Biver2019}. On a local scale, activity is driven by the surface and subsurface conditions, such as the accumulation of gas pockets and subsurface/surface stresses, the volatile ice content, morphological and topographic features that change the thermophysical condition locally (i.e., sinkholes, shadows).

The activity is able to alter the nucleus shape and to modify the internal structure. Clear morphological changes have been observed on the nucleus, related to global diffuse and local activity \citep{Elmaarry2019}. Diffuse, solar-driven activity on a global scale can explain non-gravitational forces in torque and orbit \citep{Attree2019,Kramer2019}.  Specific morphology changes seem to follow orbital cycles triggering the displacement of surface regolith \citep{2017MNRAS.469S.357K}. Other changes (i.e., landslides, fissures, sinkhole collapses, honeycombs, dunes \citep{2017Sci...355.1392E,2017A&A...604A.114H}) are more related to the local activity and subject to the diurnal variations. Some morphologic regions are deemed to be more active than others, e.g., the Anhur and Bes southern regions of 67P \citep{2017MNRAS.469S..93F}, 
due to higher erosion rates during the perihelion passage when the surface is exposed to a higher solar flux.  As a result of surface dust removal, more pristine ice-rich layers are exposed  \citep{2017MNRAS.469S..93F}. 

A major deficiency in {\it Rosetta} data is the lack of constraints on the physical properties of the surface  and subsurface layers, and the relationship of volatiles and refractories at depth. This information forms an important input to our understanding of cometary activity, and is necessary to relate coma properties, such as chemical abundances and the dust-to-gas ratio, to intrinsic nucleus properties.  Thermophysical parameters strongly depend on microphysical details such as, the material composition, the morphology and structure of the material (arrangement of the material, coordination number, void spaces) and the mixing of the different components (mineral phases, organics and ices) \citep{2019SSRv..215...19F}. One can envisage ices surrounding the dust component, ice matrices within dust matrices, and ices as a component isolated from the refractories. How minor species are mixed with water ice is unknown (e.g., trapped in amorphous ice, clathrate hydrates, pure ices), whereas this affects the conditions under which they are released. Microphysical properties affect the macroscopic behaviour of the cometary surface, such as tensile strength 
\citep{2018MNRAS.479.1273G} and compressive strength \citep{2016A&A...587A.128L}. The (somewhat disputed) high compressive strength measured from {\it Philae} just below the surface of 67P  suggests the presence of sintering processes, driven by sublimation and recondensation of water \citep{2015Sci...349b0464S}.

Values for the thermal inertia (very low, 10--50 SI, \citep{2015Sci...347a0709G,2018A&A...616A.122M}) and porosity ($>$ 70\%, \citep{2016MNRAS.462S.516H,2019MNRAS.483.2337P}) have been obtained, but the thermal conductivity and heat capacity are not fully constrained, and neither are the gas regime transport and recondensation. The refractory-to-ice ratio in the nucleus remains a subject of lively debate \citep{Chou2019}. The relative amounts of volatile species (e.g., CO, noble gases) with respect to water measured in the coma certainly do not reflect their abundance in pristine ices, since the depth of the sublimation fronts  are different. The seasonal production patterns of minor species fall in two groups, following either H$_2$O or CO$_2$, and are not correlated with sublimation temperatures \citep{2015A&A...583A...4L,2017MNRAS.469S.108G}. This can be interpreted as a consequence of two different ice phases, H$_2$O or CO$_2$ ice-dominated, in which the minor species are embedded in different relative abundances. Whether this is a primordial heterogeneity or due to evolution and differentiation is unclear. 

{\bf A primary objective for a future comet mission should be improved knowledge of the thermophysical parameters and microphysical properties of the comet surface and subsurface. Properties to be studied as a function of depth include composition, particle size distribution and shapes, morphology and bulk properties of the material, mixing of the different components, location of the sublimation fronts, ice-phase (e.g., whether amorphous or crystalline in the deep interior), and relationships between near-nucleus coma and sub-surface properties. Different mission scenarios with varying levels of complexity can be envisaged, from in situ (lander-based) measurements to a cryogenic sample return. Drilling at depths of $>$ 1 m in regions deprived of dust airfall (which is technically challenging) may permit access to  volatile-rich layers and amorphous water ice.} The detection of CO$_2$ ice at the surface of 67P shows that volatiles can condense on the surface during the winter season when they are experiencing a years-long night \citep{2016Sci...354.1563F} (Fig.~\ref{fig:activity}).


\subsection{What was the role of comets in the delivery of volatiles and prebiotic compounds to early Earth?} 
\label{sec:astrobio}

\begin{figure}
\parbox[b]{.6\linewidth}{
\includegraphics[height=5.8cm]{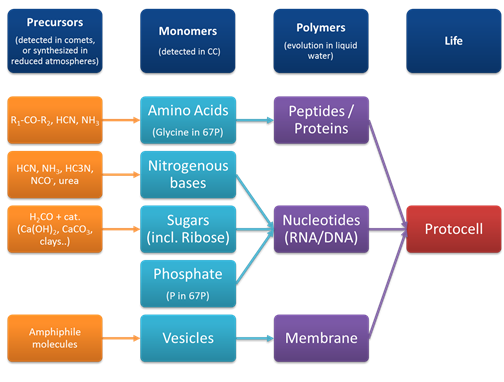}}\hfill
	\parbox[b]{.35\linewidth}{\caption{The formation of a protocell from precursors detected in comets and/or carbonaceous meteorites. Amino acids are the building block of proteins, nucleobases; ribose and phosphate are the building blocks of nucleotides (which are the building blocks of RNA and DNA), and amphiphilic molecules are known to spontaneously self-assemble into vesicles in water (i.e., into primitive cell membranes). From \citep{2017SSRv..209....1C}.\label{fig:exobio} }}
\end{figure}

Since comets are reservoirs of a large number of ingredients necessary for the origin of life, like water and organic molecules, they are targets of prime interest for any astrobiology investigation \citep{2017SSRv..209....1C,1997abos.conf...97O}. 

The link between comets and the origin of water on Earth has always been puzzling and debated. The water D/H ratio measured in comets is quite diverse: ranging from the terrestrial value in comets 46P and 103P ($\sim$1.5$\times$10$^{-4}$) \citep{2011Natur.478..218H,2019A&A...625L...5L}, to more than 3 times this number in comet 67P \citep{2015Sci...347A.387A}. No relation between the D/H of water and the dynamical family of the comet has been established so far \citep{2015Sci...347A.387A}. However, it has been recently proposed that comets may all have the same D/H in their water ice and that the variation in the observed ratio could be due to fractionation effects during sublimation  \citep{2019A&A...625L...5L}. Deciphering the origin of water on Earth requires not only measuring the D/H in water in more comets to further assess D/H variability among comets (which can be done, albeit with difficulties, from Earth-based observatories or space telescopes), but also to measure it in different phases (gas and icy particles) within a single comet  \citep{2019A&A...625L...5L}. The last point requires in-situ measurements and/or cryogenic sample return. Beyond water, the contribution of comets to the Earth budget of other volatile compounds (i.e., to the formation of the atmosphere) can also be further constrained by measuring isotopes of noble gases, such as Xe and Ar \citep{2016E&PSL.441...91M,2017Sci...356.1069M}. Xe and Ar isotopes have only been measured in comet 67P and they bring strong constraints on the budget of volatiles brought to Earth (20\% of atmospheric Xe is of cometary origin). Their relative abundances have to be confirmed beyond the case of  67P (to test their consistency in comets in general). 


For a long time, comets have been considered as a source of organic compounds which could have played a key role in the chemistry leading to the origin of life \citep{1961Natur.190..389O}. Key constituents for prebiotic chemistry have been identified (Fig.~\ref{fig:exobio}), e.g., HCN, formamide (NH$_2$CHO), 
glycolaldehyde (HCOCH$_2$OH), the simplest sugar and an important intermediate on the path towards forming more complex biologically relevant molecules \citep{2015SciA....1E0863B}, and glycine, the simplest of the amino acids, which are the building blocks of proteins \citep{2016SciA....2E0285A}. Besides organic compounds, phosphorus, a key element in the structure of nucleotides, the building blocks of DNA and RNA, has also been detected in the coma of 67P \citep{2016SciA....2E0285A}. Furthermore, ammonium salts, also detected in 67P \citep{Poch2019,Altwegg2019}, are known to be key precursors in the syntheses of amino acids and nucleobases. {\bf Further characterization of the chemical nature of the cometary organics will bring strong constraints to the list of chemical ingredients delivered to the early Earth, potentially enabling the chemical evolution that led to the origin of life on our planet, and maybe elsewhere.}

The origin of chirality is also a very important question in prebiotic chemistry. Is the asymmetry of life already `written' in the organic material brought from space, or is it the result of a stereospecific mechanism on the early Earth, or a pure random selection followed by amplification while life developed? Measurements in carbonaceous chondrites show a few percent excess of the L configuration for amino acids \citep{1997Sci...275..951C}. {\bf Chiral measurements in comets are necessary to assess the extent to which enantiomeric excesses are widely distributed among primitive small bodies of the Solar System and among families of chemical compounds within a given comet. Such observations would  support a scenario in which the Solar System was formed in a region where ice-rich particles were irradiated during the protoplanetary phase by an external source of circularly polarized light, inducing stereo-specific photochemistry \citep{2014ApJ...788...79M}. The determination of chirality will be possible by in situ or by sample return measurements} (see planned measurements from {\it Philae} \citep{2014P&SS..103..318G}).



\subsection{How do the dust coma, the surrounding plasma, and the nucleus interact? } 
\label{sec:plasma}

The expanding coma around a cometary nucleus is ionised by solar Extreme UltraViolet (EUV) radiation and energetic electrons \citep{1987AdSpR...7..147C,2018A&A...618A..77H}. This forms the cometary ionosphere, composed of free electrons and cations (e.g., H$_2$O$^+$ and H$_3$O$^+$). The cometary ionosphere is surrounded by the solar wind plasma, made of free electrons, protons and alpha particles, which interacts with the cometary coma, e.g., through charge exchange \citep{Wedlund2019}.

{\bf \underline{Dust--plasma.} How are cometary dust dynamics and properties affected by charging processes? What is the feedback on a cometary plasma?}
Comets are natural dusty plasma laboratories which evolve along their elliptical orbits due to changes in dust properties (density, size, porosity) and outgassing. In the presence of a gas-phase plasma (here, the cometary ionosphere) and close enough to the nucleus, the dust particles become negatively charged (as free electrons stick to the particle’s surface), attract gas-phase low-energy cations, and grow in size. The dust charging is expected to influence the dust dynamics and spatial distribution (through electromagnetic forces), the dust physical characteristics (e.g., erosion) and the gas-phase plasma properties (e.g., charge imbalance, electron cooling) (e.g., \citep{Mendis2013}). 
Charged nanograins were first detected at a comet during the {\it Rosetta} mission \citep{2015GeoRL..42.6575B}. They are accelerated outward by gas drag and, away from the nucleus, by the solar-wind induced electric field \citep{2015A&A...583A..23G}. Nanograins produced by dust disintegration in the coma may be responsible for the attenuation, near perihelion, of the EUV solar flux measured at {\it Rosetta} location \citep{2017MNRAS.469S.626J}. Modelling predicts that dust charging affects particles via electrostatic disruption and erosion, and depletes the plasma of electrons with respect to cations \citep{2015ApJ...798..130V}. Observation of  electron depletion may have been hindered at comet 67P due to the relatively low dust abundance and {\it Rosetta}'s very negative potential, but it has been observed in space in other dusty plasmas. Understanding  
dusty plasma properties is of interest to nuclear fusion for energy production. Though dust  has been observed in space in other dusty plasmas (protoplanetary disks \citep{2012ApJ...744....8M}, Titan's atmosphere \citep{Lavvas2013}, and Enceladus' plume \citep{2011JGRA..11612221M}) the associated dynamical properties of dusty plasma are not yet understood because of lack of dedicated observations. {\bf To visit a comet with instrumentation optimised to detect and assess dusty plasma would not only address how cometary dust interacts with ionospheric plasma but also contribute to increase our understanding of dusty plasma for the benefit of the planetary, astrophysical, and plasma communities. }

{\bf \underline{Neutral gas--plasma.} How does partially collisional plasma behave? How does it influence the large-scale structure (e.g., diamagnetic cavity)? How is it affected by transient events?}
Comets offer a unique neutral-plasma environment which evolves from collisional to collisionless regimes with changes in outgassing and cometocentric distance, similar to the partially ionized neutral-plasma environment expected in protoplanetary disks. It is critical to assess the nature of the plasma fine structure, as it plays a critical role in shaping large-scale structures. At comet 67P, the diamagnetic cavity boundary has been linked to instabilities between the unmagnetised, collisional and magnetised, collisionless plasmas \citep{2017MNRAS.469S.372H}. As the outgassing increases, ion-neutral chemistry occurs, driving ion composition \citep{2017MNRAS.469S.427H}. The collisional regime may be affected by solar (coronal mass ejections, co-rotative interactive regions) and cometary (outbursts) transient events \citep{2017A&A...607A..34H,Goetz2018}. 
While {\it Rosetta} was first to unveil the evolution of the coma and its interaction with the solar wind, observations of the plasma dynamics and the boundary location were hampered by the energy resolution of the ion sensors, {\it Rosetta}'s very negative potential, and the time resolution of the plasma and field sensors. 
Quasi-collisionless environments are found in planetary exospheres and comae, and are difficult to model as they are at the frontier between kinetic and fluid approaches. It is not clear how collisional and electromagnetic processes are linked together and which role the ion-neutral friction plays. It is hence critical to acquire observational constraints, especially in the absence of satisfying theoretical approaches. With an expanding coma, neutral densities are higher at comets for a given level of collisionality, which makes measurements easier. Comparing the nature of the diamagnetic cavity present in different levels of collisionality and magnetisation (e.g., at comets, after supernovae explosion, or in laser experiments, \citep{Winske2019}) is of high relevance to different communities.

{\bf \underline{Nucleus--plasma.} How is the nucleus affected by the plasma environment, including solar extreme events? Do interactions with the solar wind influence the activity and evolution of comets?}
At low cometary activity, solar wind sputtering on dust particles residing on the nucleus surface releases non-volatile species \citep{2015A&A...583A..22W}. Cometary ion sputtering may be a source of neutrals \citep{2017NatCo...815298Y}, though not the main one for O$_2$ \citep{Heritier2018}. Bombarded by the solar wind, the nucleus is predicted to become electrically charged; the fine dust on its surface could be electrostatically levitated and ejected away \citep{2015P&SS..119...24N}. The plasma influence on the nucleus surface during extreme solar events and over time is not yet understood, but may be vital for the assessment of cometary activity and evolution. 








\begin{sidewaystable}
\caption{{\footnotesize Traceability Matrix (part 1 of 2). See Sect.~3 and Fig.~\ref{tab:mission-types}. Type of mission$^\dag$: O=Orbiter, L=Lander, {\bf SR}=Sample Return, {\bf CSR}=Cryogenic Sample Return. Target$^\ddag$: JFC=Jupiter Family Comet, CEN=Centaur, DC=Dormant Comet, LP=Long Period comet, DN= Dynamically New Comet, MBC=Main Belt Comet. Instruments Acronyms: Atomic Force Microscope (AFM), Alpha Particle X-Ray Spectrometer (APXS), Circular Dichroism Spectroscopy (CDS), Fourier-Transform  Ion-Cyclotron Resonance (FT-ICR),Gas Chromatography (GC), High Resolution Inductively Coupled Plasma Mass Spectroscopy (HR-ICP-MS), Inductively Coupled Plasma Mass Spectrometry(ICPMS), Inductivity Coupled Plasma-Optical Emission Spectroscopy (ICP-OES), Insoluble Organic Matter (IOM), Isotopic Ratio Mass Spectroscopy (IRMS), Liquid Chromatography/Laser Desorption Mass Spectrometry (LC/LD-MS), Scanning Electron  Microscopy (SEM), Secondary Ion Mass Spectrometry (SIMS), Soluble Organic Matter (SOM), Time-Of-Flight (TOF), Transmission Spectroscopy (TEM), Thermal ionization Spectroscopy (TIMS), X-Ray Powder Diffraction (XRD), X-ray Absorption Near-Edge Structure (XANES). {\it Note 1}: to be developed from Langmuir Probe/Mutual Independence Probe.}}
\vspace{-0.5cm}
\footnotesize
\begin{center}
       { \renewcommand{\arraystretch}{1.1}
\begin{tabular}{|c|c|c|c|c|c|c|c|c|c|}
\hline
\rowcolor{Snow2}
 &  &  &  & \multicolumn{6}{|c|}{}\\
\rowcolor{Snow2}
 \bf{Scientific goal} & \bf{Measurement} & \bf{Instrument} & \bf{Mission}$^\dag$  & \multicolumn{6}{|c|}{\bf{Target$^\ddag$}} \\ \cline{5-10} 
\rowcolor{Snow2}  &  &  &  & \bf{JFC} & \bf{CEN} & \bf{DC} & \bf{LP} & \bf{DN} & \bf{MBC}  \\
\hline
\rowcolor{beaublue!50}
\multicolumn{10}{|c|}{\bf{How and where did cometary materials get assembled? Which post-planetesimal evolution paths need to be considered?}}\\
\hline
\rowcolor{LightCyan}
 Nucleus Internal structure & Internal mass distribution & Radio Science & O & X & X & X & X & X & X \\
\rowcolor{LightCyan}
                          & Deep internal structure   & Low-frequency radar & O+L,O & X & X &  & X &  & X \\
\hline 
\rowcolor{LightCyan}
Nucleus Surface Properties & Morphology & High-Res colour camera & O, L & X & X & X & X & X & X \\
\rowcolor{LightCyan}
layering			   & 3D shape model & Medium-Res camera, Altimeter & O & X & X & X & X & X & X \\	
\rowcolor{LightCyan}
&  Shallow subsurface structure, layering &	High-frequency radar & O & X & X & X & X & X & X \\					   
\rowcolor{LightCyan}                           
                           & Temperature and thermal inertia & MIR Radiometer, IR spectrometer & O, L & X & X & X & X & X & X \\
\hline
\rowcolor{LightCyan}
Comet building blocks & Pebbles, dust particles morphology & Dust analysers, AFM & O, L & X & X & X & X & X & X \\
\rowcolor{LightCyan}
& size distribution, porosity, & High Res cameras & O, L & X & X & X & X & X & X \\
& composition & Low-frequency radar & O+L, O, L & X & X & X & X & X & X \\
\hline
\rowcolor{LightCyan}
Petrologic assemblage phase & Optical properties, mineralogy, & microscopy, SEM, TEM, & L, {\bf SR} & X & & & & &  \\
\rowcolor{LightCyan}
relationships between minerals, & cristallinity, molecular& Raman, IR spectroscopy, XRD & &  & &  &  &  &  \\
\rowcolor{LightCyan}
between organics and minerals & composition of bulk and grains & & & & & & & &  \\
\hline

\rowcolor{beaublue!50}
\multicolumn{10}{|c|}{\bf{What is the presolar heritage of cometary materials? What do comets tell us about large-scale mixing and dynamical processes in the early Solar System?}} \\
\hline
\rowcolor{LightCyan}
Timescales for formation & radiometric ages from  & SIMS, TIMS, ICPMS & {\bf SR} & X & &  &  &  &  \\
\rowcolor{LightCyan}
of cometary matter & short- and long-lived radio isotopes & &  & & & & & &   \\
\hline
\rowcolor{Gold}
Composition of organic matter & Mass spectra, infrared   & TOF-SIMS, XANES & L, {\bf SR} & X & &  &  &  &  \\
\rowcolor{Gold}
up to masses of 1000 Da & spectra & spectroscopy (IOM), GC/LC-MS &  & & & & & &   \\
\rowcolor{Gold}
& & and  FT-ICR/Orbitrap (SOM) &  & & & & & &  \\
\rowcolor{Gold}
& &  MS/IR Spect. & O & X & X & & X & X & X \\
\hline 
\rowcolor{Gold}
Origin of organic matter & C, N, H isotopic & IRMS,GC-IRMS & {\bf SR}, {\bf CSR} & X & &  &  &  &  \\
\rowcolor{Gold}
& composition  &  NanoSIMS & &   & & & & &  \\
\hline
\rowcolor{Gold}
Water ice origin & D/H, $^{18}$O/$^{16}$O, $^{17}$O/$^{16}$O & IRMS, NanoSIMS, Laser Spect. & {\bf CSR} & X & &  &  &  &  \\
\rowcolor{Gold}
&  (within 0.1\% for CSR) & Mass spectrometer & O & X & X & & X & X & X \\
\rowcolor{Gold}
& Form (crystalline versus amorphous) & Thermal probes & L & X & &  &  &  &   \\
\hline
\rowcolor{Gold}
Volatile presolar heritage  & Molecular and isotopic, &  Mass Spectrometer  & {\bf CSR} & X & &  &  &  &  \\
\rowcolor{Gold}
Semi-volatiles (e.g., salts) & abundances, noble gases & Gas chromatography & O, L & X & X & & X & X & X  \\
\hline
\rowcolor{Gold}
Composition of mineral phases & Mineralogy, spectroscopy & microscopy, SEM, TEM,& {\bf SR} & X & &  &  &  &  \\
\rowcolor{Gold}
&& Raman, IR spect., & &  & &  &  &  &  \\
\rowcolor{Gold}
& & APXS, VIS-IR, mid-IR spect.& O, L & X & X & X & X & X & X \\
\hline
\rowcolor{Gold}
Elemental abundances & Intensity of mass spectral lines & HR-ICP-MS, ICP-OES & {\bf SR} & X & &  &  &  & \\
\rowcolor{Gold}
asteroids/planets/Sun relationships & & MS, Orbitrap & O, L & X & X &  & X & X & X \\
\hline
\rowcolor{Gold}
Presolar grains & Isotope ratios of major elements ,  & NanoSIMS, SIMS, IRMS, TEM & {\bf SR} & X & &  &  &  &  \\
\rowcolor{Gold}
& with 1\% accuracy, mineralogy & & &  & &  &  &  &   \\
\hline
\multicolumn{9}{r}{Continued on the next page.} \\

\end{tabular}
}
\end{center}
\label{TM1}
\end{sidewaystable}

\newpage

\begin{sidewaystable}
\caption{Traceability Matrix (part 2/2).}
\footnotesize
\begin{center}
      { \renewcommand{\arraystretch}{1.1}
\begin{tabular}{|c|c|c|c|c|c|c|c|c|c|}
\multicolumn{10}{l}{Continued from previous page.} \\
\hline
\rowcolor{Snow2}
 &  &  &  & \multicolumn{6}{|c|}{}\\
\rowcolor{Snow2}
 \bf{Scientific goal} & \bf{Measurement} & \bf{Instrument} & \bf{Mission}$^\dag$  & \multicolumn{6}{|c|}{\bf{Target$^\ddag$}} \\ \cline{5-10} 
\rowcolor{Snow2}  &  &  &  & \bf{JFC} & \bf{CEN} & \bf{DC} & \bf{LP} & \bf{DN} & \bf{MBC}  \\
\hline
\rowcolor{beaublue!50}
\multicolumn{10}{|c|}{\bf{How does comet activity work? How do surface and coma observations reconnect with the pristine, deep interior?}}\\
\hline
\rowcolor{Burlywood}                           
Nucleus subsurface/surface & Temperature and thermal inertia & MIR/mm Radiometer, IR spectrometer & O, L & X & X & X & X & X & X \\
\rowcolor{Burlywood} 
thermal properties & Temperature depth profile & Penetrometer/Thermal sensors & L & X & &  &  &  &  \\ 
Active layers properties & Dielectric properties & High-frequency Radar & O & X & X &  & X & X & X \\
\hline
\rowcolor{Burlywood}
Subsurface mechanical properties & Material strength and layering & Static/dynamic penetrometer & L & X & &  &  &  &  \\
\hline 
\rowcolor{Burlywood}
Comet activity & Coma dust and gas distributions & Colour Camera/IR spectroscopy & O & X & X & X & X & X & X \\
\rowcolor{Burlywood}
diurnal/seasonal variations & night-side activity & Dust-impact analysers & O, L & X & X & X & X & X & X\\
\rowcolor{Burlywood}
& & Mass spectrometer & O, L & X & X &  & X & X & X \\
\hline
\rowcolor{Burlywood}
Activity-induced surface & Morphology, colour & Medium Res Colour Camera & O, L & X & X & X & X & X & X \\
\rowcolor{Burlywood}
changes & composition, ice distribution & VIS-IR spectrometer & O, L & X & X & X & X & X & X \\
 & Mass transfer, erosion & High-freq radar, Altimeter & O & X & X &  & X & X & X \\
\hline
\rowcolor{beaublue!50}
\multicolumn{10}{|c|}{\bf{What was the role of comets in the delivery of volatiles and prebiotic compounds to early Earth?}}\\
\hline
\rowcolor{Pink}
Prebiotic compounds & Abundance of amino acids & LC-MS, GC-MS, & {\bf CSR}, {\bf SR}, L & X & &  &  &  &  \\
\rowcolor{Pink}
& and sugar-related compounds & LD-MS (Orbitrap) &  &  & &  &  &  &   \\
\hline
\rowcolor{Pink}
Chirality & Enantiomeric proportions & LC-MS, GC-MS, CDS & {\bf SR}, L & X & &  &  &   & \\
\hline
\rowcolor{Pink}
Earth volatiles & Noble gases abundance & MS & {\bf SR}, {\bf CSR} & X & &  &  &  &  \\
\rowcolor{Pink}
&  and isotopic composition  &  & O, L & X   &  & & X & X &  \\
\hline
\rowcolor{Pink}
Earth formation & Highly siderophile elements & ICP-MS, TIMS & {\bf SR} & X & &  &  &  &  \\
\rowcolor{Pink}
&  abundances (Os, Ir, Ru, Rh, ..) &  & &  & &  &  &  &  \\
\hline
\rowcolor{beaublue!50}
\multicolumn{10}{|c|}{\bf{How do the dust coma, the surrounding plasma, and the nucleus interact?}} \\
\hline
\rowcolor{Green}
Dust-plasma interaction & Electron and positive ion densities & see note 1 in caption & O & X &  X &  & X & X & X \\
\rowcolor{Green}
& Electric field& Mutual Impedance/Langmuir probes & & X & X &  & X & X & X \\
\rowcolor{Green}
& Negative ion density & Negative Ion Composition Analyser & & X & X &  & X & X & X\\
\rowcolor{Green}
& Dust grain flux \& size distribution  & Dust Analysers & & X & X & X & X & X & X\\
\rowcolor{Green}
& (nano to micro) & & &  & &  &  &  &  \\
\rowcolor{Green}
& EUV/FUV brightness (Sun) & Langmuir probe, UV Spect.& & X & X & & X & X & X  \\
\hline
\rowcolor{Green}
Coma-plasma interaction & Ion and neutral composition & Ion and Neutral Mass Spectrometer& O & X & X &  & X & X & X \\
\rowcolor{Green}
(collisionality) & and number density & Pressure gauge & & X & X &  & X & X & X \\
\rowcolor{Green}
& e$^{-}$ number density/temperature,  &  Mutual Impedance/Langmuir probes & & X & X &  & X & X & X \\
\rowcolor{Green}
& ion bulk velocity & & &  &  &  &  &  &  \\
\rowcolor{Green}
& Energetic electron/ion flux& $e^⁻$ \& ion composition analyser & & X & X & X & X & X & X \\
\rowcolor{Green}
& Magnetic field magnitude & Flux Gate Magnetometer & & X & X & X & X & X & X \\
\rowcolor{Green}
& \& components & & &  &  &  &  &  &  \\
\hline
\rowcolor{Green}
Nucleus-plasma interaction & Same as above box & Same as above box & O, L & X & X & X & X & X & X \\
\rowcolor{Green}
(including solar events) &  FUV emission brightness  & FUV spectrometer& & X & X & X & X & X & X \\
\rowcolor{Green}
& Fluxes of major neutral species & IR/Submm High-Res spectr. &  & X & X & X & X & X & X \\
\hline
\end{tabular}
}
\end{center}
\label{TM2}
\end{sidewaystable}


\begin{figure}[h!]
\includegraphics[width=\textwidth]{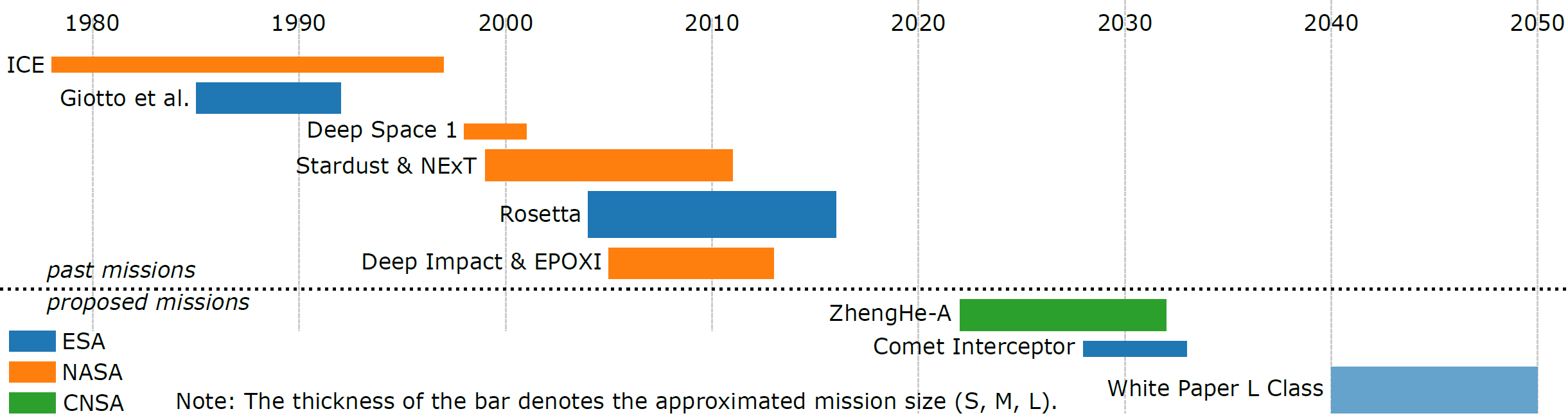}
\caption{Roadmap of comet space exploration.}
\label{fig:roadmap}
\end{figure}

\textcolor{white}{\section{{\bf Missions and technological requirements}\label{sec:Mission}}}

\begin{figure}
\parbox[b]{.6\linewidth}{
\includegraphics[width=10.5cm]{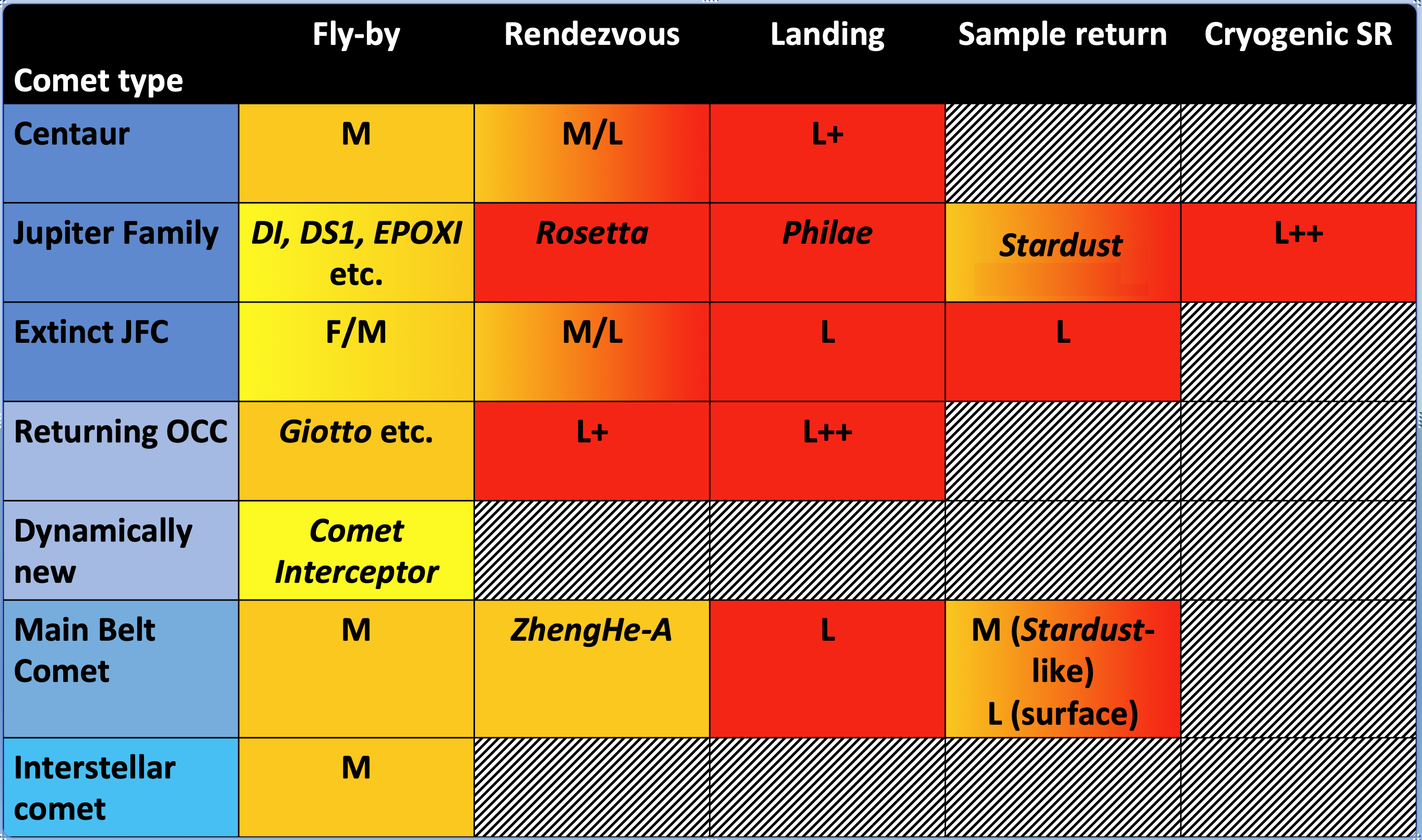}}\hfill
	\parbox[b]{.35\linewidth}{\caption{{\small Approximate mission classes for different mission and comet types, in increasing complexity from left to right, and covering different evolution stages of comets from the four possible reservoirs (Kuiper Belt, the Oort cloud, the Main Belt, and other planetary systems). Shading indicates approximate cost from yellow (F-class) through orange (M-class) to red (L-class or multi-agency flagship missions). Hatched boxes indicate that such a combination  is not seen as feasible, mostly due to excessive $\Delta$v requirements. Past and planned missions are shown.}} \label{tab:mission-types}}	
	\end{figure}

In Section 2, we have shown that a large number of major questions need to be addressed by future comet space missions in order to give final answers on the composition and evolution of these bodies and their relationships with other primitive Solar System bodies, with important implications for understanding Solar System formation and evolution, and planet habitability.

\par
For the ESA Voyage 2050 long-term (2035-2050) plan, \emph{{\bf \textcolor{blue(pigment)}{AMBITION}}} and associated technical challenges should set the bar (see the prospective for future comet missions in \citep{Thomas2019}). Also to be considered is the international  context of comet space exploration, in particular the pre-selection of the NASA New Frontiers Comet Astrobiology Exploration Sample Return ({\it CAESAR}) mission for a Phase A study (see roadmap in Fig.~\ref{fig:roadmap} and details in Sect.~\ref{sec:worldwide}). Though {\it CAESAR} was not finally selected, it is a strong candidate for the next New Frontiers call (likely to be issued in 2020). Future cometary exploration can be performed by both ESA's L and/or M-class mission budgetary envelopes. We are proposing to adopt the \textbf{L-class} mission scenario for a mission including an \textbf{Orbiting spacecraft (O)} with a \textbf{Lander (L)} (Sect.  \ref{sec:lander}) and \textbf{(Cryogenic) Sample Return (CSR-SR)} (Sect.  \ref{sct:samplereturn}). Conversely, the \textbf{M-class} mission scenario, limited to an orbiter, seems more suitable for targeting MBCs or Centaurs (Sect. \ref{sec:othertypes}).
In the Traceability Matrix Table \ref{TM1}-\ref{TM2}, we present the appropriate instrumentation (though the list is not exhaustive), including that for the analysis of  returned samples, and the associated mission type for different classes of cometary families.

\par
{\bf Returning a sample collected at depth, or, still more challenging, at cryogenic temperatures and preserving the stratigraphy of the surface layers, is the next step of comet exploration, addressing questions related to the thermophysical and microphysical properties of subsurface layers (Sects~\ref{sec:assemblage}, \ref{sec:activity}), in addition to the nature and formation of comet materials (Sects~ \ref{sec:heritage}, \ref{sec:mixing}, \ref{sec:astrobio}). Laboratory analyses of returned samples} achieve high precision, high resolution measurements {\bf that cannot be performed in space} (see e.g., \citep{2008M&PS...43..367R}).

\par
Despite the paramount importance of a cryogenic sample return approach, several key measurements (Table \ref{TM1}--\ref{TM2})  still need instruments operating on orbiter and lander platforms. 
These include the investigation of the deep interior and active layers to be performed by low-frequency (monostatic or bistatic) and high-frequency radars (Sect.~\ref{sec:assemblage}), isotopic ratios in trace volatiles (Sect.~\ref{sec:heritage}), and plasma science (Sect.~\ref{sec:plasma}). Also, the understanding of diffuse activity requires combining sub-surface thermophysical and near-surface coma measurements, necessitating a lander (Sect.~\ref{sec:activity}). A landing mission with drills and possibly mobility (or multiple static landers) is attractive in this respect. Instrumentation could include e.g., microscopic imaging, thermal and permittivity probes, and analytical capabilities such as miniaturized mass spectrometers (see Table \ref{TM1}-\ref{TM2}).  We provide details in Sect.~\ref{sec:lander} about next-generation landers for the exploration of small Solar System bodies.

To achieve safe landing and sampling, the orbital spacecraft requires at least a narrow angle camera for site selection. To achieve an optimal scientific choice of landing site and understand the context around it, the payload should also include wide angle cameras and IR spectro-imagers, along with dust-impact and gas analysers. Key surface properties such as temperature, thermal inertia, albedo, and ice content, can be measured using VIS-IR spectrometers and MIR/submm spectro/radio- meters. 
   

{\it Rosetta} explored a Jupiter-family comet (JFC) originating from the Kuiper-Belt. This population of comets has perihelia of typically 1 AU and aphelia near the orbit of Jupiter, and is accessible by rendezvous, landing and sample return missions. These comets are the most suitable targets for sample return. An attractive subset of this population are the hyperactive comets, which are releasing large amounts of icy aggregates and show a terrestrial D/H ratio in water, unlike 67P \citep{2019A&A...625L...5L}. Prototypes are comet 103P/Hartley 2, encountered by the NASA {\it EPOXI} mission, and 46P/Wirtanen, the initial {\it Rosetta} target. The mechanisms driving hyperactivity are unknown, and could be related to CO$_2$ driven-activity, as observed for 103P/Hartley 2 \citep{2011Sci...332.1396A}, to their small size \citep{2019A&A...625L...5L}, with the possibility that these comets are particularly ice-rich. A landing mission to a hyperactive comet (likely an L-class mission) would provide important insights about activity processes and comet diversity.     

Some populations of comets are still completely unexplored. We discuss in Sect.~ \ref{sec:othertypes} a variety of different sized missions towards extinct/dormant comets (DC), MBCs, long-period (LP) and dynamically new comets (DN), Centaurs (CEN) and interstellar objects (ISOs). In the Traceability Matrix Table \ref{TM1}-\ref{TM2},  we assumed rendezvous missions for DC, MBC, CEN and LP, and flyby/orbiter science for DN (appropriate also for ISOs). Approximate mission classes are indicated in Fig.~\ref{tab:mission-types}.\\




\subsection{Lander}
\label{sec:lander}

The delivery and deployment of a lander on a cometary nucleus is quite challenging due to the low gravity and cometary activity. The available mass for payloads, the generation of power and the establishment of the communication link with the orbiter are further limitation to the science which could be achieved from an autonomous laboratory housed on a surface lander.

{\it Philae} was a complex device with sophisticated mechanisms, designed to cope with a wide range of possible cometary environment and conceptual redundancy (e.g., anchoring harpoons plus 'ice-screws' and a damping mechanism in the landing gear) \citep{2008SSRv..138..275B,2009AdSpR..44..847U}.   By taking advantage of technological developments of the last 25 years, a next generation lander  could be improved in several aspects.

{\bf Miniaturization}:   Not only miniaturization in computers, but also in instrumentation,  have advanced over the last years.  This, however, does not necessarily apply for mechanisms (e.g., a drill designed to reach a certain depth can hardly be designed smaller).  Nevertheless, sensors and electronics would require less mass and volume for a future lander (assuming similar performance).  One could also consider a change in concept:  while {\it Philae} was one large lander with ten instruments and sophisticated capabilities (like drilling or rotation), other systems (like the much smaller {\it MASCOT/Hayabusa2} lander delivered to asteroid Ryugu \citep{2017SSRv..208..339H}) are more modest in their technical requirements (and variety of science that can be performed), but also more flexible to use.  Beyond {\it MASCOT} there is a trend to apply CubeSats for interplanetary missions, including landing (e.g., Juventus or APEX aboard the ESA {\it Hera} asteroid mission).

{\bf Enhanced landing system}: While {\it Philae} descended passively after being ejected from the mother spacecraft, nowadays there exist highly developed systems with 3-axis stabilization, propulsion and powerful Guidance, Navigation and Control (GNC) ready to be used for small spacecraft or landers.  This could improve the flexibility in the landing scenario and increase reliability.   Autonomous obstacle recognition and hazard avoidance would minimize landing risks due to surface roughness or boulders, and allow a very accurate touch-down with small landing uncertainties.  This could possibly even allow targeting  active spots or exposed ice patches that have been identified beforehand with orbiter instruments.

{\bf Mobility}:  In line with a 3-axis propulsion system (e.g., cold gas thrusters), mobility becomes also more easily achievable. Hopping by using the landing gear mechanism was considered for {\it Philae}, but given up in favor of safe anchoring  (ironically, the anchoring did not work, and {\it Philae} made an un-controlled hop).  {\it MASCOT} allowed some relocation, using an internal torquer.  The concept worked fine, but the hops were very short in distance ($\sim$70~cm).  For {\it MMX} (JAXA/Martian Moons eXplorer) it is foreseen to deliver a small ($\sim$30~kg) rover to Phobos, and drive several tens of meters in low gravity environment.  A new lander design may well allow the investigation of several areas on a cometary surface via mobility, thus, giving insight into the heterogeneity of surface properties and enhancing possible (radar) sounding experiments.

{\bf Instrumentation}: instrument development has improved since the 1990s and new types of instruments (e.g., Raman spectrometers, LIBS, tele-microscopes) now do have space applications.  Other instrument types (like mass spectrometers) are available with better performance/resolution.


\subsection{(Cryogenic) Sample return} 
\label{sct:samplereturn}

 \noindent
 Sample collection and return to Earth is the fourth step of robotic space exploration, after flybys, orbiter remote science and lander in-situ science.  Each step gets more challenging and usually builds up on the experiences of previous missions.
So far, samples have been returned robotically from the Moon ({\it Apollo} \& {\it Luna} missions), solar wind ({\it Genesis}), comet 81P/Wild 2 ({\it Stardust}), and asteroid Itokawa ({\it Hayabusa}).  Sample return missions to other asteroids are underway ({\it Hayabusa2} at Ryugu and {\it OSIRIS-REx} at Bennu) or in the phase of realization (Sect.~\ref{sec:worldwide}). In the past, an advanced study of the Asteroid Redirect Robotic Mission ({\it ARRM}) was completed with the aim of grabbing  a multi-ton boulder from the surface of a near-Earth object \cite{2015AcAau.117..163M}. Cometary sample return has been studied as the original concept of {\it Rosetta} \cite{Atzei1994}, as well as for {\it CAESAR}/NASA mission  returning to comet 67P \cite{2019LPI....50.2541G}. In Europe, numerous industrial studies have been performed, in particular during phase A studies at ESA of the {\it MarcoPolo} and {\it MarcoPolo-R} asteroid sample return projects \citep{2012ExA....33..645B} for the Cosmic Vision programme (2015-2025), including sampling tool technology and the re-entry capsule (e.g., heat shield material development, aerodynamic stability). The {\it Triple-F mission}, a Comet Nucleus Cryogenic Sample return, has also been proposed to the Cosmic Vision programme, in collaboration with the Russian space agency \citep{2009ExA....23..809K}.  

For a future mission, considering an advance in technology, the requirements for sampling, storage and return to Earth shall be beyond those expressed in the last NASA decadal survey (non-cryogenic, surface only), but should dare to attempt to go a step further.
Table \ref{tbl:samplereturn} lists different sampling strategies with increasing complexity.

\begin{table}[h!]
\begin{center}
\begin{tabular}{|c|c|c|}
\hline 
\rowcolor{Snow2}
\bf{Requirement} & \bf{Possible technique} & \bf{Studies with references} \\
\hline 
Coma dust & Aerogel capture & {\it Stardust}\citep{2006Sci...314.1711B} \\
\hline 
Surface dust only & Brush, air-blow system, etc	& {\it OSIRIS-REx}\citep{2017SSRv..212..925L}, {\it CAESAR}\cite{2019LPI....50.2541G} \\
\hline
Surface material, incl. consolidated 	 & Corer, combined  &	{\it CAESAR}\citep{2019LPI....50.2541G} \\
material and volatiles & brush/rock-chipper & {\it CORSAIR}-Sampler\citep{Volk2018} \\
\hline
Boulder capture & Grabber \& manipulator arm & {\it ARRM}\citep{2015AcAau.117..163M} \\
\hline
Subsurface material & Corer, & {\it Philae} drill with retrieval \\
to a depth of few cm & digging system & system, {\it CORSAIR}-Corer \\
\hline
Sub-surface core, & Complex Corer & 'original' {\it Rosetta}\citep{Atzei1994}\\
protecting the stratigraphy &  &  \\
\hline
Subsurface material, & Corer plus & 'original' {\it Rosetta}\citep{Atzei1994} \\
kept at cryogenic temperatures, & cryogenic chain & NASA Tech. Study\citep{Veverka2011} \\
during sampling, storage and return	&  & \\
\hline
\end{tabular}
\end{center}
\caption{Sample return mission requirements and collection techniques.}
\label{tbl:samplereturn}
\end{table}


{\bf The  (Cryogenic) Sample Return option is the most ambitious scenario for cometary exploration, in which a sample of surface/subsurface material is collected and returned to Earth.} A suitable protocol of cosmochemical measurements to be conducted in terrestrial laboratories to fully characterize the sample has been defined (Table~\ref{TM1}-\ref{TM2}). A further advantage of the sample return scenario is that the analysis techniques are not limited by the restricted resources available onboard the spacecraft and by the technology readiness available, but can make use of the full potentialities of terrestrial laboratories, for which future developments in analytical capabilities are possible.
\par 
In this scenario two possible options are feasible: {\bf 1) \emph{Sample Return} will provide a surface/subsurface sample with the scope to characterize its mineral and organic material phases. The sample return canister will  not be pressurized nor stabilized at cryogenic temperatures, so the volatile fraction will not be preserved during the return voyage. 2) \emph{Cryogenic Sample Return} is similar to the Sample Return option, but with the possibility to pressurize and thermally stabilize the sample at cryogenic temperatures during the entire return voyage to preserve the volatile fraction distribution.} A major issue of this second option is the definition of the cryogenic storage temperature and overpressure of the sample after the collection, because these parameters will define the survivability of the volatile species. 
\par
\begin{figure}
\parbox[b]{.4\linewidth}{
\includegraphics[width=10cm]{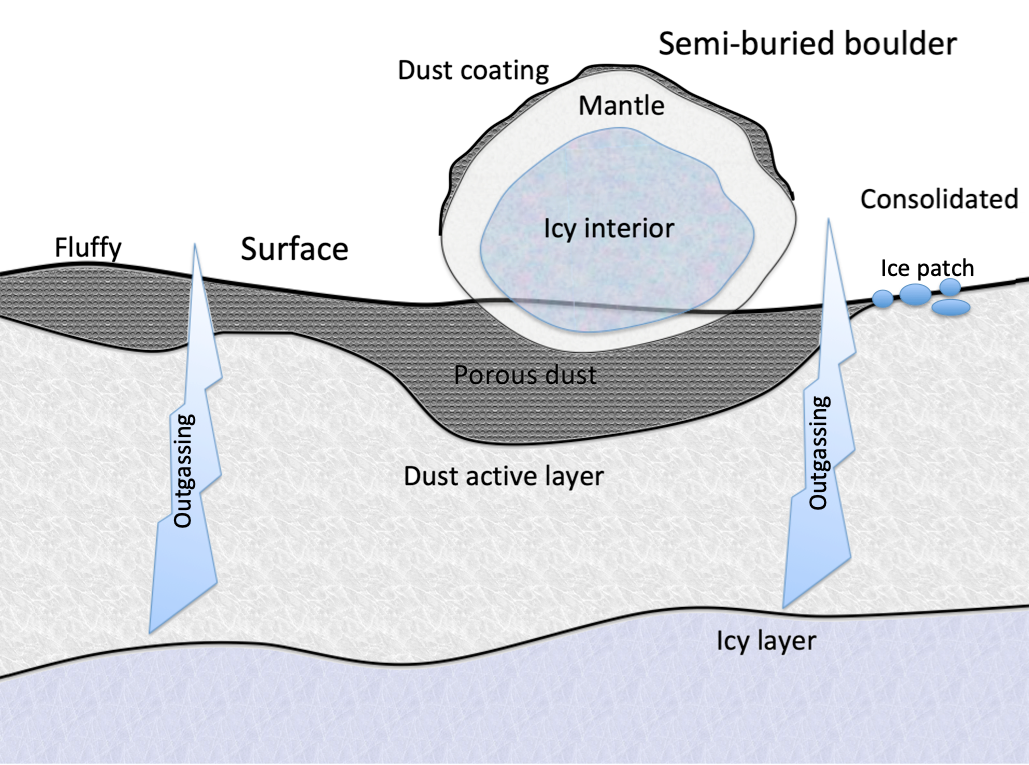}}\hfill
	\parbox[b]{.4\linewidth}{\caption{{\small Schematic diagram attempting to reconcile the information we currently have on the interior layers and surface of cometary nuclei. Adapted from Thomas et al. MIARD project report.}} 
	\label{fig:surface_scheme}}
\end{figure}

From the {\it Rosetta} and {\it Philae} measurements, it appears that 67P surface is characterized by a relatively hard sintered layer at a depth of a few centimetres, upon which sits a layer of loosely consolidated regolith (see Fig. \ref{fig:surface_scheme}). The observed presence of airfall implies that the thickness of this layer of `sand' is likely to be highly variable but prevalent above the winter hemisphere at perihelion. This material is also subject to gas-driven transport leading to the formation of dune-like structures up to 3 meters in height \citep{2017A&A...604A.114H,2017MNRAS.469S.357K}. Hence, we need to be careful about what we actually mean by `surface' because the regolith, being depleted of volatile species, may not be a representative sample, at least for some science goals. This is a dynamic environment comprising both the residues of particulate matter that have become concentrated as a consequence of the devolatilization operating in the surface layers, and the products of any alteration processes that arise from interactions of surface materials with insolation/radiation etc. It is inevitable that the chemical and physical properties of this surface layer will be different from those of the bulk comet. The nucleus's  surface can be made of dehydrated dust \cite{2015Sci...347a0628C} or of more consolidated terrains \cite{2015Sci...347a1044S}: due to periodic solar heating and the consequent outgassing of volatiles, the surface is mainly enriched by mineral and organic material but is almost depleted of ices, with the exception of very localized areas (detailed below). This is at the root of the difficulties in measuring the cometary ice-to-dust ratio, a fundamental parameter to link cometary formation conditions with interstellar grain models and observations. In order to respond to this question {\bf it is necessary to excavate the surface until ice-rich layers are reached}. 
\par
{\bf The  requirements on the cryogenic sample acquisition and storage during the return to Earth are the following}: 1) a continuous cylindrical sample from the nucleus surface down to a desirable depth of about 3 m, but not less than 1 m, with a diameter of about 10 cm (corresponding to a volume of about 23.5 liters and mass of about 12 kg); 2) the sampled material should not be mechanical nor thermal altered by the extraction and storing process to maintain stratigraphy; the sample can be stored in a series of, e.g., 6 canisters corresponding to incremental depths of 0.5 m each; 3) since the volatile fraction can be expected preferentially in the lower part of the core sample the canisters housing the inner layers need to be sealed (to maintain controlled overpressure) and thermally stabilized at cryogenic temperature ($\leq$90 K). Pressure and temperature of the canister need to be kept both under control to guarantee the survivability of the volatile species in the sample. Apart very volatile species, like CO, for which saturation vapor pressure is very high (2.4 bar), an overpressure of about 1 bar is sufficient to maintain stable H$_2$O, CO$_2$ and HCN ices at $T$ = 90~K (see Table 6.3 of \citep{Veverka2011}).

\par
While the 3-meter-depth sample will offer the optimal approach to understand the internal structure and composition of a cometary nuclei surface, another possibility to collect volatile-rich cometary samples is offered by the presence of exposed and semi-buried boulders, some up to meter-size, directly accessible from the surface (Fig. \ref{fig:surface_scheme}). The interior of compact boulders may in fact still contain high amount of water ice, as evidenced by high resolution images from {\it Rosetta}. Alternatively, several localized areas, up to a few tens of meters in size, indicate the presence of exposed water ice patches \cite{2015A&A...583A..25P, 2016A&A...595A.102B, 2016Natur.529..368F} or condensed water \cite{2015Natur.525..500D} and carbon dioxide \cite{2016Sci...354.1563F} (Fig.~\ref{fig:activity}) directly accessible on the surface, thus allowing access to condensed  volatile species (albeit with not pristine relative abundances) without the necessity to excavate the interior layers. Grabbing and preserving at controlled temperature one of those boulders or volatile-rich patches would be an alternative approach for a Cryogenic Sample Return mission, although there remain questions as to whether these are representative of the volatile composition of the deep interior. 
\par
Any sampling strategy on a cometary body faces several technological challenges in addition to those well known for deep space missions, including the development of a navigation system in the proximity of the nucleus surface, the landing and anchoring system, the uncertainty of the nucleus' surface properties (roughness, strength, compactness), the sampling/excavation system (stratigraphy preservation) and the uninterrupted cryogenic temperature and pressure controlling systems during the sampling process, handling, return to Earth, atmospheric re-entry, retrieval on Earth and during curation.

\subsection{M-class Main-Belt comet/Centaur rendezvous} 
\label{sec:othertypes}

In depth understanding of a single Jupiter-family comet that goes beyond what {\it Rosetta} achieved will require a large mission. However, there is also much to be learned by expanding our knowledge of different types of comets, or different evolutionary stages, with some populations still completely unexplored. A variety of different sized missions could potentially contribute to exploration of these populations (Fig. \ref{tab:mission-types}), from first reconnaissance flyby missions (e.g., the selected F-class {\it Comet Interceptor}) to more in depth in situ analysis with rendezvous missions. The different orbits of the various comet classes dictate the cost and complexity of these missions.

{\bf Extinct (or dormant) comets} can be found within the near-Earth object population, so a first flyby to explore this end-state of comet evolution could be achieved at relatively low cost (e.g., F-class dedicated mission, or as part of a multi-comet/asteroid tour in the M-class category, similar to the {\it CASTAway} concept proposed to the M5 call \citep{2018AdSpR..62.1998B}). More advanced missions require a better understanding of which are truly extinct comets and how these are separated from D-type asteroids perturbed into higher eccentricity orbits. Also, since such comets are inactive, landing and \mbox{(sub-)surface} sampling would be required to make in situ composition measurements, implying a large mission. Extinct comets have elliptical orbits, meaning that rendezvous or landing missions require significant $\Delta$v (a mesure of the impulse required to perform a maneuver). 

At the other end of the short-period comet evolutionary path we have the {\bf Centaurs}, objects with relatively unstable orbits in the giant planet region that are in a transition from Kuiper Belt objects to JFCs \citep{2019arXiv190508892P}. These have a wide range in orbits, from the nearly circular orbit of 29P/Schwassmann-Wachmann 1 just outside of Jupiter, to those that orbit beyond Uranus, and those with eccentric orbits that cross planet orbits. They also vary widely in size, including bodies much larger than any known comet nucleus, like Chariklo, a $\sim$200 km diameter minor planet with its own ring system \citep{2014Natur.508...72B}. Only a fraction ($< 10 \%$) of Centaurs are observed to show comet-like activity, triggered likely by the sublimation of hypervolatiles such as CO, as measured for 29P. The activity of Centaurs is often punctuated by large outbursts. For example, 29P shows almost constant activity and very regular large (many magnitudes increase in coma brightness) outbursts, the triggering mechanism of which being unknown. {\bf A mission to a Centaur would therefore be very interesting for comet science, to understand how comet activity works beyond the water ice snowline, and to see an earlier phase in comet evolution for comparison with results from JFC missions.} The challenges of such a mission are primarily due to the large distance from the Sun of the Centaurs, implying long cruise times and low power levels for a solar-powered spacecraft. Even a flyby mission would most likely be at least an M-class, and a rendezvous would certainly be at the upper end of M-class and into L-class. The scientific payload required would include remote sensing instruments to study the nucleus and in situ instrumentation to measure gas and dust composition, with the mission class and available budget dictating the choice between a minimal suite of a camera and mass spectrometer for a basic investigation, up to a {\it Rosetta}-like array of instruments for an in-depth study. The difficulty of such a mission depends greatly on how distant a target is chosen -- 29P, just outside of Jupiter, is probably the most feasible, with its circular orbit lending itself well to electric propulsion. Finally, a first look at a Centaur could be provided by a flyby en-route to Uranus or Neptune, if an Ice Giants mission is selected by any agency, provided a suitable target is found.




The {\bf Main Belt Comets} are a relatively easily accessed population, as they have low eccentricity orbits within the asteroid belt.  The {\it Castalia} MBC rendezvous mission was proposed to the M4 and M5 calls \citep{2018AdSpR..62.1947S}, with the latter version passing programmatic and technical constraints and receiving positive feedback from the panel, even if it was not selected in the end.  The goals of this mission were to make a first exploration of this new class of comets, confirm the presence of water ice in small Main Belt objects, understand how the activity of MBCs worked, test whether or not the water they contain is compatible with the Main Belt Comets being the source of Earth's water, and finally to use all of these results to better constrain models of Solar System formation and evolution. The proposed instrument payload built on {\it Rosetta} heritage to allow a direct comparison between MBCs and JFCs, and to enable the necessary very sensitive in situ measurements. The very low activity level of MBCs dictated a prolonged period at very close range to the comet during its few-month active period, but advances in spacecraft autonomy in navigation relative to {\it Rosetta}'s capabilities make such a proposal feasible as a medium sized mission. A (sub-)surface package (lander/penetrator) is not feasible for such a mission, but would certainly add further capabilities in a larger class mission, especially if it has the capability to reach volatile-rich layers, which are thought to be buried at some metres depth over most of the surface of MBCs. The recently approved  {\it ZhengHe-A} Chinese mission plans to make the first visit to an MBC (Sect. \ref{sec:worldwide}), but, as far as details are available at this time, does not expect to operate close to the comet during its active period, or sample the sub-surface, meaning that its ability to measure volatile composition will be very limited. Another proposal to the NASA Discovery call, {\it Proteus}, proposes a similar mission to {\it Castalia}, but, if not selected, an ESA MBC mission remains a compelling case for a future M-class comet mission. 


Finally, the exploration of {\bf comets from more distant reservoirs} has been very limited so far. {\it Giotto} and the other spacecraft in the `Halley armada' flew by {\bf 1P/Halley}, an evolved returning {\bf Oort cloud comet} that is now in a relatively short-period orbit, while the newly selected  F-class mission {\it Comet Interceptor} is expected to make the first flyby of a new, pristine, Oort cloud comet entering the inner Solar System for the first time. For such new comets, more advanced missions (rendezvous, etc) are unlikely to be feasible any time soon due to the short warning time between discovery and perihelion and the very large $\Delta$v that would be required to match the speed of the comet. A rendezvous with a returning comet, such as 1P on its next return in 2061, could be imagined, but would be a large mission, and the scientific gain versus a more capable mission at a short-period comet, including possible sample return, is not obvious. 

{\bf Comets from other star systems}, i.e., interstellar objects (ISOs), present a similar,  but even more extreme problem, but would be very exciting mission targets (as the only feasible way to study extrasolar material in situ). Especially, if future examples are similar to the first discovered 'Oumuamua' and display little visible activity, they will only be discovered with very short warning time \citep{2017Natur.552..378M}. It is expected that the Large Synodic Survey Telescope (LSST) will find more ISOs, perhaps 1 per year:  we will have a better idea of the true population characteristics and arrival rate by the 2030s. A mission similar to {\it Comet Interceptor}, i.e., designed to wait in space for a suitable target to be found, and then make a fast flyby, could then be imagined to encounter an ISO. This would probably be a more expensive mission than {\it Comet Interceptor}, as it would likely require significantly more $\Delta$v and a rapid reaction operations scheme, but it could be proposed within the M-class budget.

\textcolor{white}{\section{{\bf International context of comet space and Earth-based  exploration}}}

\label{sec:worldwide}

\noindent
{\bf Space exploration.} Numerous space missions to small Solar System bodies have been proposed to ESA, NASA, JAXA and other space agencies. Besides the selected ESA F-class {\it Comet Interceptor mission}, the only cometary mission in the landscape is the {\it ZhengHe-A} Chinese mission to be launched in 2022. Details of this mission are not well known, but the current plan is a sample return from the small near-Earth asteroid 2016 HO3, followed by a rendezvous with the MBC 133P/Elst-Pizarro in 2030 \citep{2019LPI....50.1045Z}. The payload of the {\it ZhengHe} orbiter is expected to include wide/narrow angle cameras, visible/near-infrared imaging and thermal emission spectrometers, dust analysers, a mass  spectrometer, $\gamma$-ray/neutron spectrometers, and a low  frequency  radar. {\it ZhengHe} also flies nano-landers for studies of surface and subsurface properties. The {\it ZhengHe} mission is technology driven and it remains unclear whether or not it will operate at 133P during its short active period, enabling comparable measurements to previous comet missions.

The Comet Astrobiology Exploration Sample Return ({\it CAESAR}) mission was pre-selected for a Phase A study in the New Frontiers NASA program, but did not pass the final selection. {\it CAESAR} was designed to acquire a minimum of 80 g of solid material from the surface of the comet, and to store sublimated volatiles  after warming in a gas containment system \citep{2019LPI....50.2642L}. {\it CAESAR}  would have arrived in 2028 at 67P, and the capsule (from JAXA) would have been delivered in 2038. 

Two sample return missions to asteroids are ongoing. After the success of the {\it Hayabusa} mission, which returned dust particles from S-type asteroid (25143) Itokawa,  JAXA launched  {\it Hayabusa2}, which is currently in orbit around C-type asteroid (162173) Ryugu. {\it Hayabusa2}, whose payload includes a lander and three small rovers, will bring surface material to Earth in 2020 \citep{2019Sci...364..268W}. The NASA New Frontiers mission {\it OSIRIS-REx} reached  B-type asteroid (101955) Bennu in December 2018, and will return the sample ($>$ 60 g of regolith) in 2023 \citep{2017SSRv..212..925L}.

 
\noindent
{\bf Earth-based observations of comets.} Comets are mostly studied by telescopes. Observations are generally focusing on the physical and chemical properties of the coma and tails. Direct studies of nucleus properties, even size and color, are difficult due to their small size and the presence of the coma, though recent progress has been made thanks to the availabilty of large tetescopes (e.g., \citep{2017MNRAS.471.2974K}). Telescopic observations of comets are unavoidable to perform statistical studies and investigate differences and links between the various comet populations, including Main-Belt comets and transition bodies such as Centaurs. Advances in comet knowledge strongly benefited from the development of new instrumentations combined with legacy  of several decades of dedicated surveys. Next generation telescopes will certainly provide breakthroughs in several aspects of comet science. For example, using the {\it James Webb Space Telescope} ({\it JWST}), it will be possible to detect the main drivers of cometary activity, H$_2$O, CO$_2$ and CO, out to unprecedented heliocentric distances, and to conceivably test the cometary nature of Main-Belt comets by the direct detection of water; spatially resolved infrared spectra will be used to study the properties of water ice particles, the nature and relative amounts of dust amorphous and crystalline silicates, and possibly detect nucleus surface signatures \citep{2016PASP..128a8009K}. Giant telescopes in the optical and near-IR, such as the  Extremely Large Telescope (ELT), also have an important potential for measuring, e.g., C, N, and H isotopic ratios in gas-phase species and detecting weakly abundant organic molecules. The  LSST is expected to vastly increase the number of known and characterized comet-like objects. In the submillimeter range, the high-resolution and sensitivity capabilities of ALMA make possible the detection of thermal emission from comet nuclei, from which insights on their size and surface thermal skin depth can be obtained; worth mentioning also is the study of gas-dust interrelations through the detailed mapping of the distributions of gases and dust particles in the coma with ALMA. {\bf Science objectives presented in this white paper cannot be addressed by telescopic observations. }

\noindent
{\bf Laboratory developments.} Laboratory experiments play a fundamental role in research programs designed to investigate the properties and evolution of comets. Experiments are diverse, and include the analyses of dust and ice analogues by complementary methods aimed at quantitative studies of morphology, structure, chemistry, and optical behavior, and simulation of processes occurring at the early stages of Solar System formation, or at the surface and in the interior of cometary nuclei, cf, e.g., the KOSI experiment \citep{1998EM&P...80..369K}. Such data aid in interpreting observations performed remotely or in situ, and provide strong constraints for theoretical models. Recently, the international network CoPhyLab (Comet Physics Laboratory), comprising scientists from TU Braunschweig, IWF Graz, University of Bern, MPS G\"{o}ttingen, DLR Berlin and the University of Stirling, was founded with the aim at experimentally and theoretically investigating the thermophysical behavior of dust-ice mixtures under cometary conditions, e.g., with various levels of insolation and with state-of-the-art dust and ice compositions and morphologies.  
Apart this collaboration, the international network of laboratories is a growing community which includes many research groups involved in the characterization of physical and chemical properties of cometary materials. \\

\textcolor{white}{\section{{\bf Conclusion}}}

This white paper proposes that \textcolor{blue(pigment)}{\it AMBITION}, a Comet Nucleus Sample Return mission, be a cornerstone of ESA's Voyage 2050 programme. Rendezvous missions  to Main Belt comets and Centaurs  are compelling cases for M-class missions, expanding our knowledge by exploring new classes of comets.   \textcolor{blue(pigment)}{\it AMBITION} would  engage  a  wide  community,  drawing expertise from a vast range of disciplines within planetary science and astrophysics. With \textcolor{blue(pigment)}{\it AMBITION}, Europe will continue its leadership in the exploration of the most primitive Solar System bodies.   



\newpage

\noindent
{\LARGE {\bf \textcolor{blue(pigment)}{References}}}
\setcitestyle{numbers}
\begingroup
\renewcommand{\section}[2]{}%
{}
\endgroup



\end{document}